%% file: finetuningMZ-FSU5.tex
\documentclass[prl,superscriptaddress,twocolumn,nofootinbib,showpacs,preprintnumbers,floatfix]{revtex4}

\usepackage{mathrsfs}
\usepackage{mathtools}
\usepackage{amsfonts,amssymb,amsmath}
\usepackage{graphicx,epsfig}
\usepackage{multirow}
\usepackage{verbatim}
\usepackage{color}
\input{rgb}

\def\fsu5{\mbox{$\cal{F}$-$SU(5)$}}
\def\bfsu5{$\boldsymbol{\mathcal{F}}$-$\boldsymbol{SU(5)}$}

\def\m1half{$M_{1/2}$}
\def\m3half{$M_{3/2}$}
\def\m32{$M_{32}$}

\def\mt2{$M_{T2}$}
\def\x2{$\chi^2$}
\def\2b{$M_{T2}b$}

\def\bs0{$B_S^0 \rightarrow \mu^+ \mu^-$}

\begin{document}

\title{Confronting Electroweak Fine-tuning with No-Scale Supergravity}

\author{Tristan Leggett}

\affiliation{George P. and Cynthia W. Mitchell Institute for Fundamental Physics and Astronomy, Texas A$\&$M University, College Station, TX 77843, USA}

\author{Tianjun Li}

\affiliation{State Key Laboratory of Theoretical Physics and Kavli Institute for Theoretical Physics China (KITPC),
Institute of Theoretical Physics, Chinese Academy of Sciences, Beijing 100190, P. R. China}

\affiliation{School of Physical Electronics, University of Electronic Science and Technology of China, 
Chengdu 610054, P. R. China }

\author{James A. Maxin}

\affiliation{Department of Physics and Engineering Physics, The University of Tulsa, Tulsa, OK 74104 USA}

\affiliation{Department of Physics and Astronomy, Ball State University, Muncie, IN 47306 USA}

\author{Dimitri V. Nanopoulos}

\affiliation{George P. and Cynthia W. Mitchell Institute for Fundamental Physics and Astronomy, Texas A$\&$M University, College Station, TX 77843, USA}

\affiliation{Astroparticle Physics Group, Houston Advanced Research Center (HARC), Mitchell Campus, Woodlands, TX 77381, USA}

\affiliation{Academy of Athens, Division of Natural Sciences, 28 Panepistimiou Avenue, Athens 10679, Greece}

\author{Joel W. Walker}

\affiliation{Department of Physics, Sam Houston State University, Huntsville, TX 77341, USA}

%%%%%%%%%%%%%%%%%%%%%%%%%%%%%%%%%%%%%%%%%%%%%%%%%%%%%%%%%%%%%%%%%%%%%%%%%%%%

\begin{abstract}

Applying No-Scale Supergravity boundary conditions at a heavy unification scale to
the Flipped $SU(5)$ grand unified theory with extra TeV-scale vector-like
multiplets, {\it i.e.} No-Scale \fsu5, we express the $Z$-boson mass $M_Z$ as an explicit function of the
boundary gaugino mass $M_{1/2}$, $M_Z^2 = M_Z^2 (M_{1/2}^2)$, with implicit dependence upon a dimensionless
ratio $c$ of the supersymmetric Higgs mixing parameter $\mu$ and $M_{1/2}$.
Setting the top Yukawa coupling consistent with $m_t = 174.3$ GeV at
$M_Z = 91.2$ GeV, the value of $c$ naturally tends toward $c \simeq 1$,
which indirectly suggests underlying action of the Giudice-Masiero mechanism.
Proportional dependence of all model scales upon the unified gaugino mass $M_{1/2}$ in the No-Scale \fsu5 model suggests one possible mechanism of confronting the electroweak fine tuning problem.
\end{abstract}

%%%%%%%%%%%%%%%%%%%%%%%%%%%%%%%%%%%%%%%%%%%%%%%%%%%%%%%%%%%%%%%%%%%%%%%%%%%%


\preprint{ACT-9-14}

\maketitle

%%%%%%%%%%%%%%%%%%%%%%%%%%%%%%%%%%%%%%%%%%%%%%%%%%%%%%%%%%%%%%%%%%%%%%%%%%%%

\section{Introduction}

Supersymmetry (SUSY) represents a solution to the {\it big} gauge hierarchy problem, logarithmically
sequestering the reference to ultra-heavy (Grand Unification, Planck, String) scales of new physics.
However, there is a residual {\it little} hierarchy problem, implicit in the gap separating
TeV-scale collider bounds on (strong production of) yet elusive colored superpartner fields
from the observation of a 125~GeV light CP-even Higgs; indeed, this same heaviness of the SUSY
scale appears equally requisite to sufficient elevation of loop contributions to the physical Higgs mass itself.
One possible mechanism for reconcilation of these considerations without an unnatural invocation of fine tuning, {\it vis-\`a-vis} unmotivated cancellation of more than (say) a few parts {\it per centum} between contributions to
physics at the electroweak (EW) scale, could be the providence of a unified framework wherein
the entire physical spectrum (Standard Model + SUSY) may be expressed as functions of a single parameter.

The SUSY framework naturally provides for interplay between quartic and quadratic field strength terms
in the scalar potential of the type essential to spontaneous destabilization of the null vacuum, the former emerging
with dimensionless gauge-squared coupling coefficients from the $D$-term, and the latter with dimensionful mass-squared
coefficients referencing the bilinear Higgs mixing scale $\mu$ from the chiral $F$-term.
Crucially though, this radiative electroweak symmetry breaking (EWSB) event, as driven by largeness of the top-quark Yukawa coupling,
is not realizable without the supplementary inclusion of soft mass terms $m_{H_{u,d}}$ and the analog $B_\mu$ of $\mu$,
which herald first the breaking of SUSY itself.  In a supergravity (SUGRA) context, these terms may be expected to appear in proportion
to the order parameter of SUSY breaking in the visible sector, as gravitationally suppressed from higher scale effects in an
appropriately configured hidden sector, namely the gravitino mass $M_{3/2}$.  The gravitino mass may itself be exponentially
radiatively suppressed relative to the high scale, plausibly and naturally taking a value in the TeV range.  The
Giudice-Masiero (GM) mechanism may be invoked to address the parallel ``$\mu$ problem'', suggesting that this SUSY-preserving
coupling may likewise be of the same order, and likewise generated as a consequence of SUSY breaking, as evaluated at the high scale.

Minimization of the Higgs scalar potential with respect to the $H_u$ and $H_d$ field directions
yields two conditions on the pair of resulting vacuum expectation values (VEVs) $(v_u,v_d)$.  The overall scale
$(v_u^2+v_d^2)^{1/2}$, in product with the gauge coefficients $(g_{\rm L}^2+{g'}_{\rm Y}^2)^{1/2}/2$, is 
usually traded for the physical $Z$-boson mass $M_Z$, whereas the relative VEV strengths are parameterized
$(\tan \beta \equiv v_u/v_d)$ by an angle $\beta$.  This allows one to solve for $\mu$ and $B_\mu$ at the
electroweak scale in terms of $M_Z$, $\tan \beta$, and the soft masses $m_{H_{u,d}}$.  When addressing the
question of fine tuning, the solution for $\mu^2$ is typically inverted as follows in Eq.~(\ref{eq:ewmin}), and an argument is
made regarding the permissible fraction of cancellation between terms on the right-hand side, whose
individual scales may substantially exceed $M_Z^2$. 
\begin{eqnarray}
\frac{M_Z^2}{2} =
\frac{m_{H_d}^2  -
\tan^2\beta ~m_{H_u}^2}{\tan^2\beta -1} -\mu^2
\label{eq:ewmin}
\end{eqnarray}

\noindent For moderately large $\tan\beta$, Eq.~(\ref{eq:ewmin}) reduces to
\begin{eqnarray}
\frac{M_Z^2}{2} \simeq -m_{H_u}^2 -\mu^2 \;.
\label{eq:EWMINL}
\end{eqnarray}

At the outset, it must be noted that some level of cancellation should here be permitted, and indeed expected, within
Eq.~(\ref{eq:EWMINL}) as an unavoidably natural consequence of the EWSB event itself.  Specifically, destabilization of the symmetric
vacuum occurs in conjunction with (the falsely tachyonic) negation of the quadratic $H_u^2$ field coefficient,
provided by $m^2_{H_u} + \mu^2 \simeq - M_Z^2/2$, which is dynamically driven to vanish and then flow negative
under the renormalization group.  However, the proverbial devil may lurk in the details, an element
of which are loop-level radiative corrections $V_{\rm eff} \Rightarrow V_{\rm tree} + V_{\rm 1-loop}$ to the effective scalar potential,
and likewise via derivatives of $V_{\rm 1-loop}$ to the minimization condition expressed in Eq.~(\ref{eq:ewmin}),
which is recast with $m^2_{H_u} \Rightarrow m^2_{H_u} + \Sigma_u^u$, and $m^2_{H_d} \Rightarrow m^2_{H_d} + \Sigma_d^d$.

\section{Various Measures of Fine-Tuning}

Several approaches to quantifying the amount of fine tuning implicit in Eq.~(\ref{eq:ewmin}) have been suggested,
one of the oldest being that $\Delta_{\rm EENZ}$~\cite{Ellis:1986yg, Barbieri:1987fn} first prescribed some 30 years ago by
Ellis, Enqvist, Nanopoulos, and Zwirner (EENZ), consisting of the maximal logarithmic $M_Z$ derivative with respect
to all fundamental parameters $\varphi_i$, evaluating at some high unification scale $\Lambda$ as is fitting for gravity-mediated SUSY breaking.
In this treatment, low fine-tuning mandates that heavy mass scales only weakly influence $M_Z$, whereas strongly correlated
scales should be light.
\begin{eqnarray}
\Delta_{\rm EENZ} = {\rm Max}\,\left\{\,
\left|
\frac{\partial\,{\rm ln}(M_Z^n)}{\partial\, {\rm ln}(\varphi_i^n)}
\right|\,
\equiv\,
\left|
\frac{\varphi_i}{M_Z}
\frac{\partial\,M_Z}{\partial\,\varphi_i}
\right|\,
\right\}_{\Lambda}
\label{eq:eenz}
\end{eqnarray}

Lately, a prescription $\Delta_{\rm EW}$ emphasizing evaluation directly at the electroweak scale~\cite{Baer:2012up,Baer:2013ava} has attracted
attention, isolating contributions to the (loop-modified) right-hand side of
Eq.~(\ref{eq:ewmin}) as a ratio with the left.
\begin{eqnarray}
&\Delta_{\rm EW}& = {\rm Max}\,\bigg\{\,
\bigg| \frac{m^2_{H_d}}{\tan^2 \beta - 1} \bigg|\,,\,
\bigg| \frac{\Sigma^d_d}{\tan^2 \beta - 1} \bigg|\,,\,
\nonumber \\
&& \bigg| \frac{m^2_{H_u}\tan^2\beta}{\tan^2 \beta - 1} \bigg|\,,\,
\bigg| \frac{\Sigma^u_u\tan^2\beta}{\tan^2 \beta - 1} \bigg|\,,\,
\mu^2\, \bigg\}_{\rm EW} \!\!\!\!\!\!\div \frac{M_Z^2}{2}
\label{eq:ew}
\end{eqnarray}

\noindent More strictly~\cite{Baer:2012up}, individual contributions to $\Sigma^{u,d}_{u,d}$ may be compared to $M_Z^2/2$ in isolation, without
allowance for cancellation even within the internal summation.

A variation~\cite{Baer:2012mv,Baer:2013ava} of the prior, dubbed $\Delta_{\rm HS}$, would further split the electroweak evaluation of the soft masses and $\mu$-term
into the sum of their corresponding values at the high scale (HS) $\Lambda$ plus a logarithmic correction from the renormalization group running,
specifically \mbox{$m^2_{H_{u,d}} \Rightarrow m^2_{H_{u,d}}(\Lambda) +  \delta m^2_{H_{u,d}}$}, and \mbox{$\mu^2 \Rightarrow \mu^2(\Lambda) + \delta\mu^2$}.
\begin{eqnarray}
&\Delta_{\rm HS}& = {\rm Max}\,\bigg\{\,
\bigg| \frac{m^2_{H_d}}{\tan^2 \beta - 1} \bigg|\,,\,
\bigg| \frac{\delta m^2_{H_d}}{\tan^2 \beta - 1} \bigg|\,,\,
\nonumber \\
&& \hspace{-32pt}
\bigg| \frac{m^2_{H_u}\tan^2\beta}{\tan^2 \beta - 1} \bigg|\,,\,
\bigg| \frac{\delta m^2_{H_u}\tan^2\beta}{\tan^2 \beta - 1} \bigg|\,,\,
\mu^2\,,\, \delta \mu^2\, \bigg\}_{\Lambda} \!\!\!\div \frac{M_Z^2}{2}
\label{eq:hs}
\end{eqnarray}

\noindent The scale $\Lambda$ referenced in Eq.~(\ref{eq:hs}) is not necessarily of GUT or Planck order, but may rather contextually refer to a SUSY breaking
messenger as low as some tens of TeV~\cite{Kitano:2006gv}, although this somewhat blurs distinction from the $\Sigma^{u,d}_{u,d}$ terms comprising Eq.~{\ref{eq:ew}}.
In any event, leading top-stop loops generate corrections \mbox{$\delta m^2_{H_u} \sim m^2_{{\tilde{t}}_1} \times \ln (m_{{\tilde{t}}_1}/\Lambda)$} that have been
used~\cite{Kitano:2006gv,Baer:2012up} to infer naturalness bounds on the light stop mass (thereby also limiting vital parallel contributions to
the Higgs mass), and similarly on the gluino mass.  It has been suggested~\cite{Kitano:2005wc,Kitano:2006gv,Baer:2012up}
that a large negative trilinear soft term $A_t$, which may be generated radiatively at the intermediate scale, may engender cancellations in
the light stop ${\tilde{t}}_1$ tuning contribution while simultaneously lifting $m_h$.

\section{Impact of Dynamics on Tuning}

In the SUGRA context, $M_Z^2$ is generically bound to dimensionful inputs $\varphi_i$ at the high scale $\Lambda$
via a bilinear functional, as shown following.  The parameters $\varphi_i$ may include scalar and gaugino
soft SUSY breaking masses (whether universal or not), the bi- and tri-linear soft terms $B_\mu$ and $A_i$, as well as the $\mu$-term.
The coefficients $C_i$ and $C_{ij}$ are calculable, in principle, under the renormalization group dynamics.
\begin{eqnarray}
M_Z^2 = \sum_i C_i \varphi_i^2(\Lambda) + \sum_{ij} C_{ij} \varphi_i(\Lambda) \varphi_j(\Lambda)
\label{eq:bilinear}
\end{eqnarray}

\noindent Applying the Eq.~(\ref{eq:eenz}) prescription, a typical contribution to the fine tuning takes the subsequent form.
\begin{eqnarray}
\frac{\partial\,{\rm ln}(M_Z^2)}{\partial\, {\rm ln}(\varphi_i)}
= \frac{\varphi_i}{M_Z^2} \times \bigg\{\, 2\,C_i\varphi_i + \sum_{j} C_{ij} \varphi_j \,\bigg\}
\label{eq:eenzsum}
\end{eqnarray}

\noindent
Comparing with Eq.~(\ref{eq:bilinear}), each individual term in the Eq.~(\ref{eq:eenzsum}) sum is observed,
modulo a possible factor of 2, to be simply the ratio of one contribution to the unified $M_Z^2$ mass, divided by $M_Z^2/2$.
The structural similarity of the $\Delta_{\rm EENZ}$ and $\Delta_{\rm HS}$ prescriptions is therefore clear.

Differing conclusions drawn with respect to the supposed naturalness of a given SUSY model construction by the various
fine tuning measures described are implicit within underlying assumptions that each makes regarding what may constitute a natural cancellation. 
For example, Eq.~(\ref{eq:eenz}) permits two types of dynamic
cancellation that are not recognized by Eq.~(\ref{eq:hs}).  The first of these is between multiple Eq.~(\ref{eq:bilinear})
terms in which a single parameter $\varphi_i$ may appear. The second is between each high-scale parameter and its
running correction, for example between $\mu^2(\Lambda)$ and $\delta\mu^2$, as represented by, and absorbed into, the
numerical coefficients $C_i$ and $C_{ij}$.  $\Delta_{\rm HS}$ is therefore a harsher metric of tuning than $\Delta_{\rm EENZ}$.
It is sometimes argued~\cite{Baer:2012cf,Baer:2013ava} that $\Delta_{\rm EW}$, which adopts a perfectly {\it a priori} view
of only the low energy particle spectrum, sets an absolute lower bound on fine tuning via holistic recombination
of high scale mass parameters and their potentially large running logarithms.  Proponents of this opinion acknowledge~\cite{Baer:2013ava}
that the suggested bound is only as strong as an assumption that the SUSY breaking soft masses $m^2_{H_{u,d}}$ and the SUSY preserving
$\mu$-term harken from wholly disparate origins, such that cancellations amongst the two classes are {\it inherently} unnatural.

The positions of the authors in the current work are that $(i)$ precisely the prior mode of cancellation is {\it made} natural by its essential
role in promoting the electroweak symmetry breaking event, and that $(ii)$ solid theoretical motivation exists
(GM mechanism) for suspecting the SUSY preserving $\mu$ scale to have yet likewise been forged in the supersymmetry breaking event.
If the former precept is granted, then the limit offered by $\Delta_{\rm EW}$ is potentially spurious.
If the latter precept is additionally granted, then the very notion of electroweak fine-tuning may be moot.

\section{Tuning in No-Scale \bfsu5}

Bottom up support for the prior opinions is provided here by consideration of a specific model of low energy physics
named \fsu5 (see Ref.~\cite{Li:2013naa} and references therein), which combines $(i)$ field content of the 
Flipped $SU(5)$ grand unified theory (GUT), with $(ii)$ a pair of hypothetical TeV-scale vector-like supermultiplets (``flippons'')
of mass $M_V$ derivable within local F-Theory model building, and $(iii)$ the boundary conditions of No-Scale Supergravity (SUGRA).
The latter amount to vanishing of the scalar soft masses and the tri-/bi-linear soft couplings $M_0 = A_0 = B_{\mu} = 0$
at the ultimate \fsu5 gauge unification scale $M_{\cal F} \simeq M_{\rm Pl}$, and are enforced dynamically by invocation of a minimal K\"ahler
potential~\cite{Ellis:1984bm,Cremmer:1983bf}.  Non-zero boundary values may be applied solely to the universal gaugino mass
$M_{1/2}$, as necessarily implied by SUSY breaking, and to the $\mu$-parameter.  In this perspective, the value of $\mu$ is
actually that evolved {\it up} from the scale dynamically established in EWSB; its retrospective similarity to (and proportional
scaling with) $M_{1/2}$ is interpreted as a {\it deeply suggestive} accident.  The model is highly constrained by the
need to likewise dynamically tether (via the Renormalization Group Equations (RGEs)) the value of $B_\mu$ generated in EWSB to its mandated vanishing at $M_{\cal F}$;
in fact, this releases the constraint typically exhausted by the fixing of $B_\mu$ to instead determine $\tan \beta \simeq 20$,
which incidentally supports the approximation adopted by Eq.~(\ref{eq:EWMINL}),
while not being so large that impact of the bottom quark Yukawa coupling is substantially heightened.

The consequence (at fixed $Z$-boson $M_Z$ and top-quark $m_t$ masses)
is an effectively ${\it one}$-${\it parameter}$ model, with all leading dynamics established by just the single degree of freedom allocated to $M_{1/2}$.
Inclusion of the flippon multiplets provides a vital modification to the $\beta$-function RGE coefficients, most notably resulting in
nullification of the color-charge running $(b_3 = 0)$ at the first loop; however, dynamic dependence on the mass scale $M_V$ is quite weak,
affecting gauge unification only via logarithmic feedback from a threshold correction term.  Some vestigial freedom
remains for the preservation of $B_\mu(M_{\cal F}) = 0$ by compensating adjustments to $\tan \beta$ and $M_V$ at fixed $M_{1/2}$,
at the price of disrupting the natural tendency of this model to supply a suitable thermal dark matter (over 99\% Bino) candidate;
to be precise, there are associated fluctuations induced in the lightest neutralino (Bino) vs. the next to the LSP (stau) mass gap that
synchronously affect the dark matter coannihilation rate.

Numerical analysis of the parameter interdependencies in No-Scale \fsu5 is conducted with {\tt SuSpect~2.34}~\cite{Djouadi:2002ze},
utilizing a proprietary codebase modification that incorporates the flippon-enhanced RGEs.
Applying the $\Delta_{\rm EW}$ measure of Eq.~(\ref{eq:ew}) to the \fsu5 model is thereby
found to indicate a level of tuning that may indeed be considered large.  Down-type contributions to Eq.~(\ref{eq:ew}) are $\tan\beta$-suppressed,
and $\Sigma^u_u$ is computed to be rather small, possibly reflecting the presence of large, negative trilinear couplings.  Dominant 
contributions to $M_Z^2/2$ are thus restricted to solely the pair of terms $m^2_{H_u}$ and $\mu^2$ appearing directly in Eq.~(\ref{eq:EWMINL}).
Each term, or its absolute square root evaluated at the EWSB scale, is larger than and roughly proportional to the boundary value of $M_{1/2}$,
with a ratio around 1.8 for $M_{1/2} \sim 400$~GeV that drops to about 1.3 for $M_{1/2} \sim 1500$~GeV.  The corresponding contributions
to $\Delta_{\rm EW}$ therefore increase from about 140 to about 1000 over the same range of inputs.  One interpretation of this circumstance,
taking $\vert m_{H_u} \vert $ and $\vert \mu \vert$ to arise from disparate mechanisms, is that narrow tracking and cancellation of the two terms
indicates fine tuning.  However, we make the case here for a very different point of view: that it is {\it precisely} the close tracking of
$\vert m_{H_u} \vert $ and $\vert \mu \vert$, as {\it dynamically} induced by electroweak symmetry breaking, and {\it preserved} under 
projection onto the boundary scale $M_{\cal F}$ by the renormalization group, in the {\it context} of a single parameter construction,
that suggests a {\it natural} underlying interdependence. 

It is not the purpose of this work to present a functioning hidden sector that is capable of producing at
some high scale the requisite SUSY breaking, along with the associated soft term and $\mu$-parameter boundary values.
It is the purpose of this work to investigate, primarily by numerical means~\footnote{A parallel analytical treatment, making directly
formal, albeit approximate, application of the No-Scale \fsu5 renormalization group and boundary conditions is underway,
with results intended to follow in a separate publication.}, the dependency of all contextual low energy
physics upon the single input scale $M_{1/2}$ (and by extension its dependence in turn upon $M_{3/2}$).  Insofar as this may be demonstrated, the considered model can successfully attack the fine-tuning problem, {\it i.e.} no large cancellations enforced coincidentally, without a common dynamic origin.  Insofar
as the relation $\mu(M_{\cal F}) \simeq M_{1/2}$ may be validated, the providence of an underlying GM mechanism is supported
through identification of the ``fingerprints'' which it has impressed upon the low energy phenomenology.

The most remarkable element of this construction may be the capacity to so severely constrain
freedom of input parameterization (with all concomitant benefits to the interrelated questions of the
little hierarchy and the $\mu$-problem) while retaining consistency (even at the purely thermal level) with dark matter
observations, limits on rare processes, and collider bounds.

\section{Gaugino Parameterization of \bfsu5}

The EW fine-tuning was numerically computed for No-Scale \fsu5 according to the Eq.~(\ref{eq:eenz}) prescription
in Ref.~\cite{Leggett:2014mza}, yielding result of ${\cal O}(1)$.  This absence of fine-tuning is equivalent to
a statement that the $Z$-boson mass $M_Z$ can be predicted in \fsu5 as a parameterized function of $M_{1/2}$; clarifying
and rationalizing this intuition in a more quantitative manner is a key intention of the present section.
First, we define a dimensionless ratio $c$ of the supersymmetric Higgs mixing parameter $\mu$ at the
unification scale $M_{\cal F}$ with the gaugino mass $M_{1/2}$.
\begin{eqnarray}
c = \frac{\mu(M_{\cal F})}{M_{1/2}}
\label{eq:c}
\end{eqnarray}

\noindent This parameter $c$ is a fixed constant if the $\mu$ term is generated via the Giudice-Masiero
mechanism~\cite{Giudice:1988yz}, which can, in principle, be computed from string theory.  Its invocation implicitly
addresses the need to otherwise explicitly refer to the $\mu$ parameter as an independent high scale input.
We shall numerically scan over arbitrary values of this parameter, although the No-Scale \fsu5 construction
will be demonstrated to prefer a narrow range near $c \simeq 1$.
The vector-like flippon mass parameter $M_V$ is expected to develop an exponential compensation (for $B_\mu = 0$ and all else constant)
\begin{eqnarray}
M_V \sim A(\lambda)e^{M_{1/2}/R(\lambda)}
\label{eq:mvexp}
\end{eqnarray}

\noindent of the fundamental scale $M_{1/2}$ due to its previously described appearance within a logarithmic threshold correction,
where the undetermined dimensionful parameters $A(\lambda)$ and $R(\lambda)$ may be sensitive (among other things)
to the top quark Yukawa coupling $\lambda$.  In the same vein, sensitivity to $\tan \beta$ is weak, and it is expected
that any residual dependencies upon either parameter within the region of interest may be Taylor-expanded for absorption
into a generic quadratic function of $M_{1/2}$.

We thus adopt an {\it ans\"atz}
\begin{eqnarray}
M_Z^2 = f_1 + f_2\,M_{1/2} + f_3\,M_{1/2}^2
\label{eq:mzquad}
\end{eqnarray}

\noindent consistent with Eq.~(\ref{eq:bilinear}), where the undetermined coefficients $f_i$ represent implicit functions
of dimensionless quantities including $c$ and $\lambda$.  Some evidence suggests that the dimensionful coefficients
$f_1$ and $f_2$ may additionally be sensitive to $B_\mu$, particularly to any potential deviations from the null
No-Scale boundary value.  If $(f_1 \lll M_{1/2}^2)$ and $(f_2 \ll M_{1/2})$, then a linearized approximation of the prior is applicable:
\begin{eqnarray}
M_Z = f_a + f_b\, M_{1/2}~~.
\label{eq:mzlinear}
\end{eqnarray}

The form of  Eq.~(\ref{eq:mzquad}) must now be verified with explicit RGE calculations.   This is accomplished
via a numerical sampling, wherein the $Z$-boson mass is floated within
$20 \leq M_Z \leq 500$ GeV, and the top quark mass (equivalently its Yukawa coupling) within $125 \leq m_t \leq 225$ GeV.
The region scanned for the gaugino mass boundary is within $100 \leq M_{1/2} \leq 1500$ GeV.
In order to truncate the scanning dimension, $M_V$ and $\tan \beta$ are explicitly parameterized functions of $M_{1/2}$
(consistent with the prior description) such that the physical region of the model
space corresponding to $M_Z = 91.2$ GeV and $m_t = 174.3$, along with a valid thermal relic density, is continuously intersected\footnote{
The No-Scale \fsu5 model space favors a top quark mass of $m_t = 174.3 - 174.4$
GeV in order to compute a Higgs boson mass of $m_h \sim 125$
GeV~\cite{:2012gk,:2012gu,Aaltonen:2012qt}.  The central world average 
top quark mass has recently ticked upward (along with an increase in precision) to $m_t = 174.34$ GeV~\cite{TEWG:2014cka},
affirming this preference.}; this may be considered equivalent to fixing the top quark Yukawa coupling (and associated higher-order feedback) within just this subordinate parameterization.
The range of the ratio $c$ from Eq.~(\ref{eq:c}) is an output of this analysis, which is run from the EWSB scale up to $M_{\cal F}$ under the RGEs.

\begin{figure}[htp]
        \centering
        \includegraphics[width=0.45\textwidth]{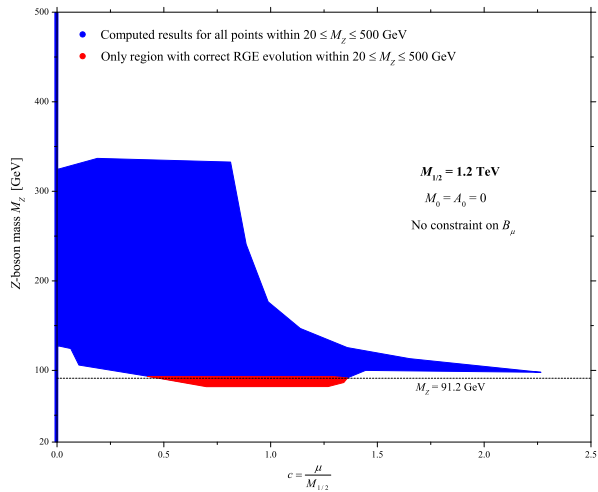}
        \caption{Representation of the region of the model space for $M_{1/2} = 1.2$ TeV that cannot
sustain correct RGE evolution (blue) versus the region that can compute a SUSY spectrum with no RGE errors,
EWSB, convergent $\mu$ term, and no tachyonic masses (red). 
All other $M_{1/2}$ produce identical results.}
        \label{fig:ewsb1200}
\end{figure}

As an initial phase of the analysis, the only constraints applied are the No-Scale SUGRA boundary conditions
$M_0 = A_0 = 0$, along with correct EWSB, a convergent $\mu$ term, and no tachyonic sfermion or pseudoscalar
Higgs boson masses, these latter conditions being the required minimum for proper RGE evolution.
These constructionist elements carve out a narrow viable parameter space between $83 \lesssim M_Z \lesssim 93$ GeV,
which is illustrated as a function of the dimensionless parameter $c$ in FIG.  \ref{fig:ewsb1200}. The blue region represents
those points that cannot satisfy the minimum evolution requirements.  The band of points clustered on the $M_Z$ axis
with $c = 0$ have no RGE solution for $\mu$ at all.  The red sliver within $83 \lesssim M_Z \lesssim 93$ GeV, narrowly
bounding the physical region around $M_Z \sim 91$ GeV, represents the sole surviving region of model space satisfying
the boundary condition $M_0 = A_0 = 0$ with a convergent RGE solution.  The gauge couplings being held fixed, the fluctuation
of $M_Z$ is attributable to fluctuation of the Higgs VEV magnitude, as compensated by differential contributions
to the right-hand side of Eq.~(\ref{eq:ewmin}) induced by variation of $\mu$, $M_{1/2}$, and the top quark Yukawa $\lambda$.

\begin{figure}[htp]
        \centering
        \includegraphics[width=0.45\textwidth]{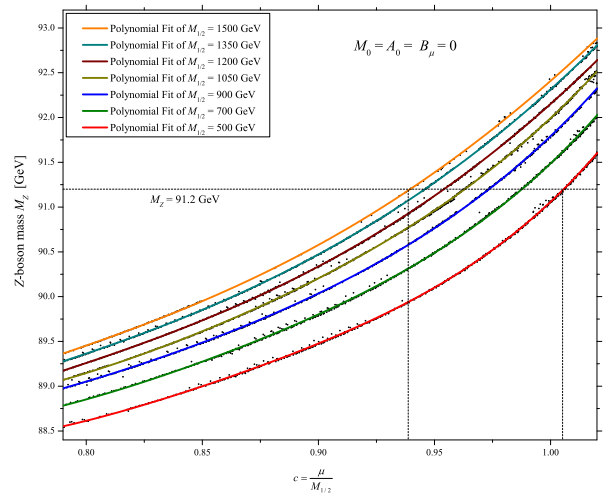}
        \caption{The $Z$-boson mass is shown as a function of the dimensionless parameter $c$ for seven
different values of $M_{1/2}$. The black points are the results of the RGE calculations, while the
curves are polynomial fits. The curves are only comprised of points with a vanishing $B_{\mu}$ parameter at the $M_{\cal F}$ unification scale.}
        \label{fig:seven_curves_c}
\end{figure}

As a second phase of the analysis, the final No-Scale SUGRA constraint $B_{\mu}
= 0$ must be applied at the $M_{\cal F}$ unification scale. The vanishing $B_{\mu} = 0$ requirement is
enforced numerically with a width $|B_{\mu}| \leq 1$ GeV that is comparable to the scale of EW radiative corrections.
The effect is to carve out a simply connected string of points from the narrow red region, depicted in FIG.
\ref{fig:seven_curves_c} for seven values of $M_{1/2}$ in the model space.

\begin{table}[htbp!]
\centering
\footnotesize
\caption{Results of RGE calculations for four different values of $M_{1/2}$. These are only points with a
vanishing $B_{\mu}$ parameter at the $M_{\cal F}$ unification scale. The entries highlighted in red are those that
compute the observed experimental measurements for $M_Z$, $m_t$, and $m_h$.}
\begin{tabular}{|c|c|c|c|c|}\cline{1-5}
\multicolumn{5}{|c|}{$M_{1/2} = 500~ {\rm GeV}$} \\ \hline
$c$&$M_Z$&$m_t$&$m_h$&$m_{\widetilde{g}}$ \\ \hline
$	~~0.955~~	$&$	~~90.189~~	$&$	~~164.47~~	$&$	~~118.88~~	$&$	~~655~~	$ \\ \hline
$	0.965	$&$	90.349	$&$	166.29	$&$	120.03	$&$	659	$ \\ \hline
$	0.975	$&$	90.543	$&$	168.34	$&$	121.30	$&$	664	$ \\ \hline
$	0.985	$&$	90.720	$&$	170.11	$&$	122.48	$&$	669	$ \\ \hline
${\color{red} 	1.005}	$&${\color{red} 	91.181}	$&${\color{red} 	174.31}	$&${\color{red} 	125.28}	$&$	{\color{red} 681	}$ \\ \hline
$	1.015	$&$	91.444	$&$	176.49	$&$	126.84	$&$	687	$ \\ \hline
$	1.025	$&$	91.741	$&$	178.81	$&$	128.45	$&$	694	$ \\ \hline
$	1.035	$&$	92.082	$&$	181.31	$&$	130.21	$&$	702	$ \\ \hline\hline
\multicolumn{5}{|c|}{$M_{1/2} = 900~ {\rm GeV}$} \\ \hline
$c$&$M_Z$&$m_t$&$m_h$&$m_{\widetilde{g}}$ \\ \hline
$	0.933	$&$	90.496	$&$	167.69	$&$	119.46	$&$	1182	$ \\ \hline
$	0.943	$&$	90.659	$&$	169.38	$&$	120.70	$&$	1190	$ \\ \hline
$	0.953	$&$	90.818	$&$	170.95	$&$	121.85	$&$	1197	$ \\ \hline
$	0.963	$&$	91.011	$&$	172.76	$&$	123.20	$&$	1205	$ \\ \hline
${\color{red} 	0.973}	$&${\color{red} 	91.189}	$&${\color{red} 	174.35}	$&${\color{red} 	124.39}	$&${\color{red} 	1212}	$ \\ \hline
$	0.983	$&$	91.390	$&$	176.06	$&$	125.71	$&$	1220	$ \\ \hline
$	0.993	$&$	91.652	$&$	178.17	$&$	127.28	$&$	1230	$ \\ \hline
$	1.003	$&$	91.849	$&$	179.67	$&$	128.53	$&$	1238	$ \\ \hline \hline
\multicolumn{5}{|c|}{$M_{1/2} = 1200~ {\rm GeV}$} \\ \hline
$c$&$M_Z$&$m_t$&$m_h$&$m_{\widetilde{g}}$ \\ \hline
$	0.913	$&$	90.525	$&$	167.96	$&$	120.31	$&$	1574	$ \\ \hline
$	0.923	$&$	90.672	$&$	169.49	$&$	121.48	$&$	1583	$ \\ \hline
$	0.933	$&$	90.832	$&$	171.06	$&$	122.66	$&$	1591	$ \\ \hline
$	0.943	$&$	90.999	$&$	172.63	$&$	123.88	$&$	1601	$ \\ \hline
${\color{red}	0.953}	$&${\color{red}	91.180}	$&${\color{red}	174.24}	$&${\color{red}	125.09}	$&${\color{red}	1610}	$ \\ \hline
$	0.963	$&$	91.371	$&$	175.87	$&$	126.41	$&$	1619	$ \\ \hline
$	0.970	$&$	91.502	$&$	176.94	$&$	127.26	$&$	1626	$ \\ \hline
$	0.983	$&$	91.780	$&$	179.11	$&$	129.04	$&$	1640	$ \\ \hline \hline
\multicolumn{5}{|c|}{$M_{1/2} = 1350~ {\rm GeV}$} \\ \hline
$c$&$M_Z$&$m_t$&$m_h$&$m_{\widetilde{g}}$ \\ \hline
$	0.905	$&$	90.537	$&$	168.06	$&$	120.76	$&$	1773	$ \\ \hline
$	0.915	$&$	90.691	$&$	169.66	$&$	121.92	$&$	1782	$ \\ \hline
$	0.925	$&$	90.850	$&$	171.21	$&$	123.12	$&$	1792	$ \\ \hline
$	0.935	$&$	91.017	$&$	172.77	$&$	124.35	$&$	1802	$ \\ \hline
${\color{red}	0.945}	$&${\color{red}	91.188}	$&${\color{red}	174.29}	$&${\color{red}	125.55}	$&${\color{red}	1811}	$ \\ \hline
$	0.955	$&$	91.367	$&$	175.81	$&$	126.77	$&$	1822	$ \\ \hline
$	0.969	$&$	91.644	$&$	178.03	$&$	128.59	$&$	1838	$ \\ \hline
$	0.977	$&$	91.806	$&$	179.28	$&$	129.62	$&$	1847	$ \\ \hline

\end{tabular}
\label{tab:results}
\end{table}

\begin{figure}[htp]
        \centering
        \includegraphics[width=0.45\textwidth]{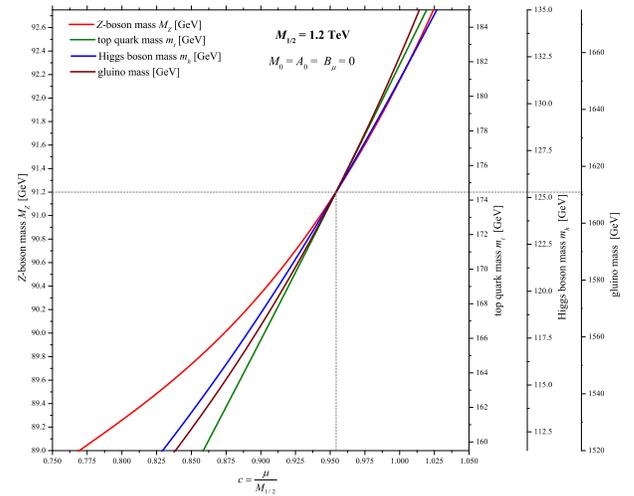}
        \caption{Depiction of the correlation between the $Z$-boson mass $M_Z$, top quark mass $m_t$, Higgs
boson mass $m_h$, and gluino mass $m_{\widetilde{g}}$, as a function of $c$, for $M_{1/2} = 1.2$ TeV.
All other SUSY particles can be expressed similarly. The curves are only comprised of points with a
vanishing $B_{\mu}$ parameter at the $M_{\cal F}$ unification scale. All other $M_{1/2}$ produce comparable correlations.}
        \label{fig:spectrum}
\end{figure}

The No-Scale SUGRA constraint on the $B_{\mu}$ parameter naturally parameterizes all the particle and
sparticle masses as a function of the dimensionless parameter $c$ of Eq.~(\ref{eq:c}). This is clearly
shown in FIG. \ref{fig:spectrum} for the $Z$-boson mass $M_Z$, top quark mass $m_t$, Higgs boson mass
$m_h$, and gluino mass $m_{\widetilde{g}}$. We use the gluino mass as an example, though the entire SUSY
spectrum can also thusly be parameterized as a function of $c$ via the $B_{\mu} = 0$ condition. The point
chosen in FIG. \ref{fig:spectrum} to exhibit the correlation between the particle and sparticle masses
is $M_{1/2} = 1200$ GeV.  TABLE \ref{tab:results} itemizes certain numerical results from the RGE
calculations for four of the FIG. \ref{fig:seven_curves_c} curves. The Higgs boson mass $m_h$ in
TABLE~\ref{tab:results} includes both the tree level+1-loop+2-loop+3-loop+4-loop
contributions~\cite{Li:2013naa} and the additional flippon contribution~\cite{Li:2011ab}.
Sensitivity is observed to fluctuation of the VEV scale with $M_Z$.

\begin{figure}[htp]
        \centering
        \includegraphics[width=0.45\textwidth]{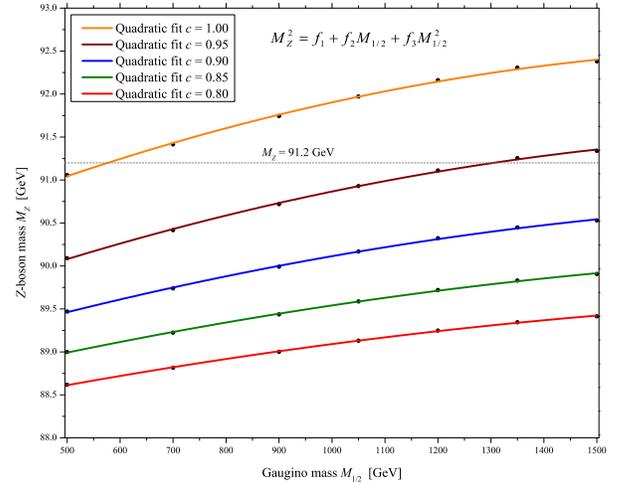}
        \caption{Simple quadratic fits for $M_Z^2$ as a function of $M_{1/2}^2$. Five different cases of $c$ are shown.
The curves are only comprised of points with a vanishing $B_{\mu}$ parameter at the $M_{\cal F}$ unification scale.
The black points are sampled from FIG. \ref{fig:seven_curves_c}.}
        \label{fig:quadratic}
\end{figure}

\begin{figure}[htp]
        \centering
        \includegraphics[width=0.45\textwidth]{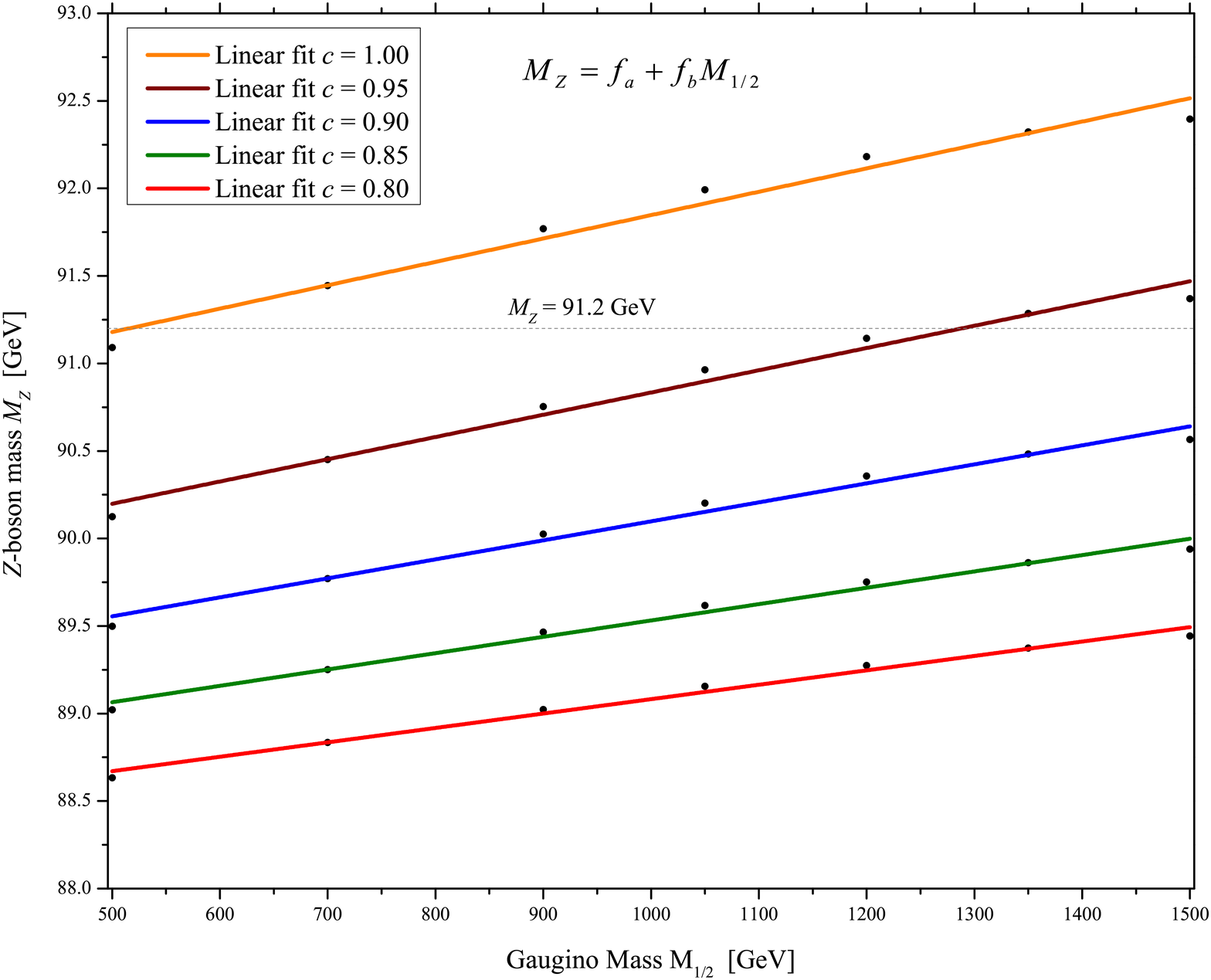}
        \caption{Linear fits for $M_Z$ as a function of $M_{1/2}$. Five different cases of $c$ are shown.
The curves are only comprised of points with a vanishing $B_{\mu}$ parameter at the $M_{\cal F}$ unification scale.
The black points are sampled from FIG. \ref{fig:seven_curves_c}.}
        \label{fig:linear}
\end{figure}

The dimensionless parameter $c$ is expected to be a fixed constant if the $\mu$ term is generated by the
GM mechanism.  On any single-valued slice of the present parameterization with respect to $c$,
particle and sparticle masses will residually be dependent upon just $M_{1/2}$,
as is visible for variation of the $Z$-boson mass in FIG. \ref{fig:seven_curves_c}.
This is made explicit for $c = 0.80, 0.85, 0.90, 0.95, 1.00$ in FIG. \ref{fig:quadratic}, where each curve is well
fit by a quadratic in the form of Eq.~(\ref{eq:mzquad}).  FIG.~\ref{fig:linear} demonstrates a fit against the linear
approximation in Eq.~(\ref{eq:mzlinear}).  As the $c$ parameter decreases, FIG.~\ref{fig:linear} illustrates that
the linear fit approaches the precision of the quadratic fit.  The dimensionful intercept $f_a$ is a function of $c$,
but is observed generically to take a value in the vicinity of $89$~GeV.  As seen in FIG.~\ref{fig:seven_curves_c}, larger
values of $M_{1/2}$ correlate with smaller values of $c$ at fixed $Z$-boson mass; it is
the region $M_{1/2} \gtrsim 900$ GeV that remains viable for probing 
a prospective SUSY signal at the 13--14 TeV LHC in 2015--16.

\begin{figure}[htp]
        \centering
        \includegraphics[width=0.45\textwidth]{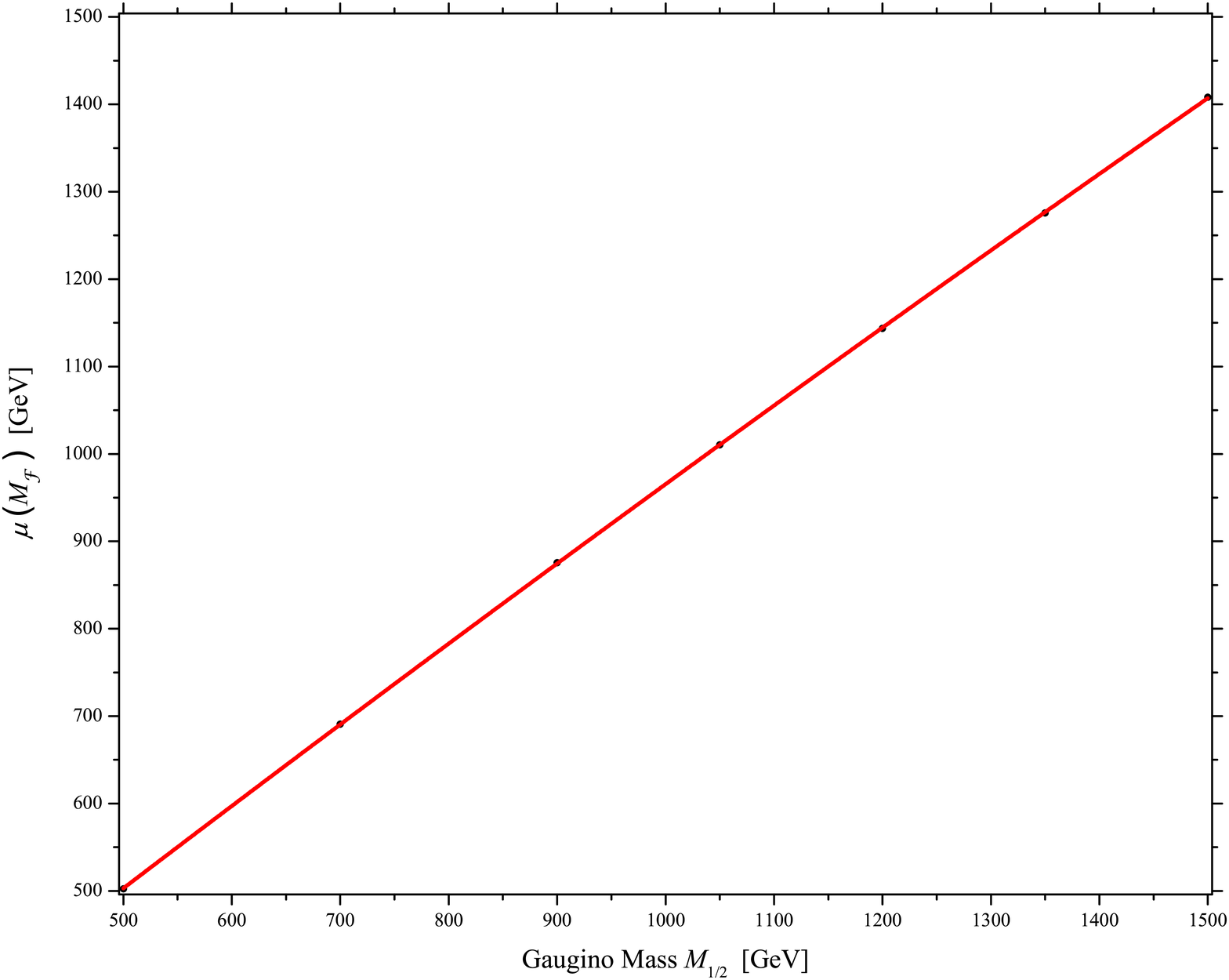}
        \caption{Linear relationship between the $\mu$ term at the $M_{\cal F}$ unification scale and $M_{1/2}$ for $M_Z =91.2$ GeV.}
        \label{fig:mu}
\end{figure}

\begin{figure}[htp]
        \centering
        \includegraphics[width=0.45\textwidth]{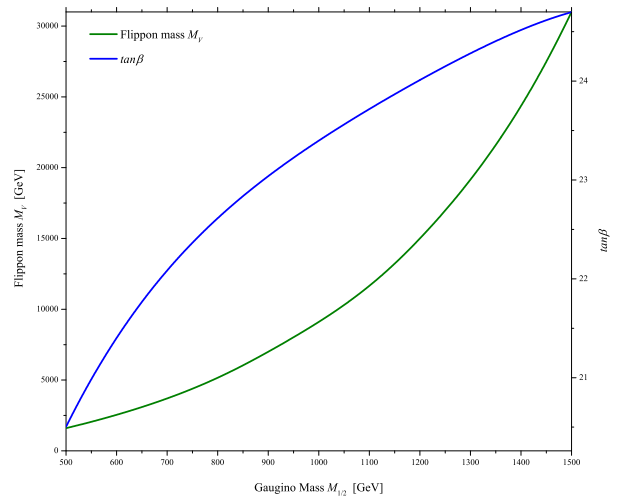}
        \caption{Vector-like flippon mass $M_V$ and $\tan\beta$ as functions of $M_{1/2}$ with Yukawa couplings fixed.}
        \label{fig:flippontanb}
\end{figure}

The relationship between the $\mu$ term and $M_{1/2}$ at the $M_{\cal F}$ unification scale is linear
for fixed $M_Z$, with a slope given by the ratio $c$ from Eq.~(\ref{eq:c}).  This is expanded in 
FIG. \ref{fig:mu} for $M_Z = 91.2$. 
Parameterization of the flippon mass $M_V$ and $\tan\beta$ as functions of $M_{1/2}$ (with the
top quark Yukawa and approximate relic density fixed) are illustrated in FIG.~\ref{fig:flippontanb}.

Having established a (family in $c$ of) quadratic expression(s) for $M_Z^2$ in the Eq.~(\ref{eq:bilinear}) form, the $Z$-boson mass is 
extracted by reference only to $M_{1/2}$ and $c$ at the high scale $\Lambda$, and fine tuning may be evaluated. 
Adopting the linear Eq.~(\ref{eq:mzlinear}) form, we first consider tuning with respect to $M_{1/2}$ at fixed $c$,
as prescribed by Eq.~(\ref{eq:eenz}).
\begin{eqnarray}
\Delta_{M_{1/2}}^{\rm \Lambda} &=& \left| \frac{\partial{\rm ln}(M_Z)}{\partial {\rm ln}(M_{1/2})} \right|
= \left| \frac{M_{1/2}}{M_Z} \frac{\partial M_Z}{\partial M_{1/2}} \right|  \\ \nonumber 
&=& \frac{1}{M_Z} \left( \frac{M_Z - f_a}{f_b} \right) f_b = 1 - \frac{f_a}{M_Z} \simeq 1 - \frac{89}{M_Z} \\ \nonumber
\label{eq:fnpar}
\end{eqnarray}

\noindent Curiously, this expression evaluates very close to zero.  It would appear this result is a consequence
of the fact that the physical $Z$-boson mass can in fact be stably realized for a large continuum of $M_{1/2}$ values,
at the expense of variation in the ratio $c$.  It may be better understood
by attending in turn to the parallel functional dependence on the dimensionless $c$ parameter itself.  We have
\begin{eqnarray}
\Delta_{c}^{\rm \Lambda} &=& \left| \frac{\partial{\rm ln}(M_Z)}{\partial {\rm ln}(c)} \right|
= \left| \frac{c}{M_Z} \frac{\partial M_Z}{\partial c} \right| \\ \nonumber
&=& \left| \frac{c}{M_Z}  \left( \frac{\partial f_a}{\partial c} + \frac{\partial f_b}{\partial c} M_{1/2} \right) \right| \sim c \simeq 1~,~\ \\ \nonumber
&&
\label{eq:fnc}
\end{eqnarray}

\noindent using the numerical observation \mbox{$\frac{\partial f_a}{\partial c} + \frac{\partial f_b}{\partial c} M_{1/2} \sim M_Z$}.
Therefore, stipulating the adopted high-scale context, we suggest that the more natural fine-tuning measure for No-Scale \fsu5 may be $\Delta_{\rm EENZ} \sim 1$.

\section{Conclusions}

We have shown here that by implementing only the No-Scale Supergravity boundary conditions of
$M_0 = A_0 = B_{\mu} = 0$ at the unification scale, the $Z$-boson mass $M_Z$ can be expressed as a simple quadratic
function of the unified gaugino mass $M_{1/2}$, $i.e.$ $M_Z^2 = M_Z^2(M_{1/2}^2)$, in the supersymmetric GUT
model No-Scale \fsu5. A top-down string theoretic construction may be expected to fix the Yukawa couplings and a 
dimensionless boundary ratio $c$ of the supersymmetric Higgs mixing parameter $\mu$ with $M_{1/2}$
at some heavy unification scale.  The only degree of freedom left to influence
$M_Z$ is $M_{1/2}$. Setting the top Yukawa coupling consistent with $m_t = 174.3$ GeV at
$M_Z = 91.2$ GeV, the value of $c$ naturally tends toward $c \simeq 1$,
which suggests underlying action of the Giudice-Masiero mechanism. The regions
of the model space in correspondence with the physical masses $M_Z = 91.2$ GeV and $m_t = 174.3$ GeV are further
consistent with the correct Higgs boson mass $m_h \simeq 125$ GeV and dark matter observations,
and possess overlap with the limits on rare processes and collider bounds.
Proportional dependence of all model scales upon the unified gaugino mass $M_{1/2}$ in the No-Scale \fsu5 model could suggest one potential mechanism of confronting the electroweak fine tuning problem.

%%%%%%%%%%%%%%%%%%%%%%%%%%%%%%%%%%%%%%%%%%%%%%%%%%%%%%%%%%%%%%%%%%%%%%%%%%%%

\section{Acknowledgements}

DVN would like to thank Andriana Paraskevopoulou for inspiration and discussions during the writing of this paper. This research was supported in part by: the DOE grant DE-FG03-95-Er-40917 (DVN), the Natural Science
Foundation of China under grant numbers 10821504, 11075194, 11135003, and 11275246, the National Basic
Research Program of China (973 Program) under grant number 2010CB833000 (TL), the Ball State
University ASPiRE Research Grant Program (JAM), and the Sam Houston State University
Enhancement Research Grant program (JWW).

%%%%%%%%%%%%%%%%%%%%%%%%%%%%%%%%%%%%%%%%%%%%%%%%%%%%%%%%%%%%%%%%%%%%%%%%%%%%

\bibliography{bibliography}

\end{document}

%% file: rgb.tex
\definecolor{AliceBlue}{rgb}{0.94,0.97,1.00}
\definecolor{AntiqueWhite1}{rgb}{1.00,0.94,0.86}
\definecolor{AntiqueWhite2}{rgb}{0.93,0.87,0.80}
\definecolor{AntiqueWhite3}{rgb}{0.80,0.75,0.69}
\definecolor{AntiqueWhite4}{rgb}{0.55,0.51,0.47}
\definecolor{AntiqueWhite}{rgb}{0.98,0.92,0.84}
\definecolor{BlanchedAlmond}{rgb}{1.00,0.92,0.80}
\definecolor{BlueViolet}{rgb}{0.54,0.17,0.89}
\definecolor{CadetBlue1}{rgb}{0.60,0.96,1.00}
\definecolor{CadetBlue2}{rgb}{0.56,0.90,0.93}
\definecolor{CadetBlue3}{rgb}{0.48,0.77,0.80}
\definecolor{CadetBlue4}{rgb}{0.33,0.53,0.55}
\definecolor{CadetBlue}{rgb}{0.37,0.62,0.63}
\definecolor{CornflowerBlue}{rgb}{0.39,0.58,0.93}
\definecolor{DarkBlue}{rgb}{0.00,0.00,0.55}
\definecolor{DarkCyan}{rgb}{0.00,0.55,0.55}
\definecolor{DarkGoldenrod1}{rgb}{1.00,0.73,0.06}
\definecolor{DarkGoldenrod2}{rgb}{0.93,0.68,0.05}
\definecolor{DarkGoldenrod3}{rgb}{0.80,0.58,0.05}
\definecolor{DarkGoldenrod4}{rgb}{0.55,0.40,0.03}
\definecolor{DarkGoldenrod}{rgb}{0.72,0.53,0.04}
\definecolor{DarkGray}{rgb}{0.66,0.66,0.66}
\definecolor{DarkGreen}{rgb}{0.00,0.39,0.00}
\definecolor{DarkGrey}{rgb}{0.66,0.66,0.66}
\definecolor{DarkKhaki}{rgb}{0.74,0.72,0.42}
\definecolor{DarkMagenta}{rgb}{0.55,0.00,0.55}
\definecolor{DarkOliveGreen1}{rgb}{0.79,1.00,0.44}
\definecolor{DarkOliveGreen2}{rgb}{0.74,0.93,0.41}
\definecolor{DarkOliveGreen3}{rgb}{0.64,0.80,0.35}
\definecolor{DarkOliveGreen4}{rgb}{0.43,0.55,0.24}
\definecolor{DarkOliveGreen}{rgb}{0.33,0.42,0.18}
\definecolor{DarkOrange1}{rgb}{1.00,0.50,0.00}
\definecolor{DarkOrange2}{rgb}{0.93,0.46,0.00}
\definecolor{DarkOrange3}{rgb}{0.80,0.40,0.00}
\definecolor{DarkOrange4}{rgb}{0.55,0.27,0.00}
\definecolor{DarkOrange}{rgb}{1.00,0.55,0.00}
\definecolor{DarkOrchid1}{rgb}{0.75,0.24,1.00}
\definecolor{DarkOrchid2}{rgb}{0.70,0.23,0.93}
\definecolor{DarkOrchid3}{rgb}{0.60,0.20,0.80}
\definecolor{DarkOrchid4}{rgb}{0.41,0.13,0.55}
\definecolor{DarkOrchid}{rgb}{0.60,0.20,0.80}
\definecolor{DarkRed}{rgb}{0.55,0.00,0.00}
\definecolor{DarkSalmon}{rgb}{0.91,0.59,0.48}
\definecolor{DarkSeaGreen1}{rgb}{0.76,1.00,0.76}
\definecolor{DarkSeaGreen2}{rgb}{0.71,0.93,0.71}
\definecolor{DarkSeaGreen3}{rgb}{0.61,0.80,0.61}
\definecolor{DarkSeaGreen4}{rgb}{0.41,0.55,0.41}
\definecolor{DarkSeaGreen}{rgb}{0.56,0.74,0.56}
\definecolor{DarkSlateBlue}{rgb}{0.28,0.24,0.55}
\definecolor{DarkSlateGray1}{rgb}{0.59,1.00,1.00}
\definecolor{DarkSlateGray2}{rgb}{0.55,0.93,0.93}
\definecolor{DarkSlateGray3}{rgb}{0.47,0.80,0.80}
\definecolor{DarkSlateGray4}{rgb}{0.32,0.55,0.55}
\definecolor{DarkSlateGray}{rgb}{0.18,0.31,0.31}
\definecolor{DarkSlateGrey}{rgb}{0.18,0.31,0.31}
\definecolor{DarkTurquoise}{rgb}{0.00,0.81,0.82}
\definecolor{DarkViolet}{rgb}{0.58,0.00,0.83}
\definecolor{DeepPink1}{rgb}{1.00,0.08,0.58}
\definecolor{DeepPink2}{rgb}{0.93,0.07,0.54}
\definecolor{DeepPink3}{rgb}{0.80,0.06,0.46}
\definecolor{DeepPink4}{rgb}{0.55,0.04,0.31}
\definecolor{DeepPink}{rgb}{1.00,0.08,0.58}
\definecolor{DeepSkyBlue1}{rgb}{0.00,0.75,1.00}
\definecolor{DeepSkyBlue2}{rgb}{0.00,0.70,0.93}
\definecolor{DeepSkyBlue3}{rgb}{0.00,0.60,0.80}
\definecolor{DeepSkyBlue4}{rgb}{0.00,0.41,0.55}
\definecolor{DeepSkyBlue}{rgb}{0.00,0.75,1.00}
\definecolor{DimGray}{rgb}{0.41,0.41,0.41}
\definecolor{DimGrey}{rgb}{0.41,0.41,0.41}
\definecolor{DodgerBlue1}{rgb}{0.12,0.56,1.00}
\definecolor{DodgerBlue2}{rgb}{0.11,0.53,0.93}
\definecolor{DodgerBlue3}{rgb}{0.09,0.45,0.80}
\definecolor{DodgerBlue4}{rgb}{0.06,0.31,0.55}
\definecolor{DodgerBlue}{rgb}{0.12,0.56,1.00}
\definecolor{FloralWhite}{rgb}{1.00,0.98,0.94}
\definecolor{ForestGreen}{rgb}{0.13,0.55,0.13}
\definecolor{GhostWhite}{rgb}{0.97,0.97,1.00}
\definecolor{GreenYellow}{rgb}{0.68,1.00,0.18}
\definecolor{HotPink1}{rgb}{1.00,0.43,0.71}
\definecolor{HotPink2}{rgb}{0.93,0.42,0.65}
\definecolor{HotPink3}{rgb}{0.80,0.38,0.56}
\definecolor{HotPink4}{rgb}{0.55,0.23,0.38}
\definecolor{HotPink}{rgb}{1.00,0.41,0.71}
\definecolor{IndianRed1}{rgb}{1.00,0.42,0.42}
\definecolor{IndianRed2}{rgb}{0.93,0.39,0.39}
\definecolor{IndianRed3}{rgb}{0.80,0.33,0.33}
\definecolor{IndianRed4}{rgb}{0.55,0.23,0.23}
\definecolor{IndianRed}{rgb}{0.80,0.36,0.36}
\definecolor{LavenderBlush1}{rgb}{1.00,0.94,0.96}
\definecolor{LavenderBlush2}{rgb}{0.93,0.88,0.90}
\definecolor{LavenderBlush3}{rgb}{0.80,0.76,0.77}
\definecolor{LavenderBlush4}{rgb}{0.55,0.51,0.53}
\definecolor{LavenderBlush}{rgb}{1.00,0.94,0.96}
\definecolor{LawnGreen}{rgb}{0.49,0.99,0.00}
\definecolor{LemonChiffon1}{rgb}{1.00,0.98,0.80}
\definecolor{LemonChiffon2}{rgb}{0.93,0.91,0.75}
\definecolor{LemonChiffon3}{rgb}{0.80,0.79,0.65}
\definecolor{LemonChiffon4}{rgb}{0.55,0.54,0.44}
\definecolor{LemonChiffon}{rgb}{1.00,0.98,0.80}
\definecolor{LightBlue1}{rgb}{0.75,0.94,1.00}
\definecolor{LightBlue2}{rgb}{0.70,0.87,0.93}
\definecolor{LightBlue3}{rgb}{0.60,0.75,0.80}
\definecolor{LightBlue4}{rgb}{0.41,0.51,0.55}
\definecolor{LightBlue}{rgb}{0.68,0.85,0.90}
\definecolor{LightCoral}{rgb}{0.94,0.50,0.50}
\definecolor{LightCyan1}{rgb}{0.88,1.00,1.00}
\definecolor{LightCyan2}{rgb}{0.82,0.93,0.93}
\definecolor{LightCyan3}{rgb}{0.71,0.80,0.80}
\definecolor{LightCyan4}{rgb}{0.48,0.55,0.55}
\definecolor{LightCyan}{rgb}{0.88,1.00,1.00}
\definecolor{LightGoldenrod1}{rgb}{1.00,0.93,0.55}
\definecolor{LightGoldenrod2}{rgb}{0.93,0.86,0.51}
\definecolor{LightGoldenrod3}{rgb}{0.80,0.75,0.44}
\definecolor{LightGoldenrod4}{rgb}{0.55,0.51,0.30}
\definecolor{LightGoldenrodYellow}{rgb}{0.98,0.98,0.82}
\definecolor{LightGoldenrod}{rgb}{0.93,0.87,0.51}
\definecolor{LightGray}{rgb}{0.83,0.83,0.83}
\definecolor{LightGreen}{rgb}{0.56,0.93,0.56}
\definecolor{LightGrey}{rgb}{0.83,0.83,0.83}
\definecolor{LightPink1}{rgb}{1.00,0.68,0.73}
\definecolor{LightPink2}{rgb}{0.93,0.64,0.68}
\definecolor{LightPink3}{rgb}{0.80,0.55,0.58}
\definecolor{LightPink4}{rgb}{0.55,0.37,0.40}
\definecolor{LightPink}{rgb}{1.00,0.71,0.76}
\definecolor{LightSalmon1}{rgb}{1.00,0.63,0.48}
\definecolor{LightSalmon2}{rgb}{0.93,0.58,0.45}
\definecolor{LightSalmon3}{rgb}{0.80,0.51,0.38}
\definecolor{LightSalmon4}{rgb}{0.55,0.34,0.26}
\definecolor{LightSalmon}{rgb}{1.00,0.63,0.48}
\definecolor{LightSeaGreen}{rgb}{0.13,0.70,0.67}
\definecolor{LightSkyBlue1}{rgb}{0.69,0.89,1.00}
\definecolor{LightSkyBlue2}{rgb}{0.64,0.83,0.93}
\definecolor{LightSkyBlue3}{rgb}{0.55,0.71,0.80}
\definecolor{LightSkyBlue4}{rgb}{0.38,0.48,0.55}
\definecolor{LightSkyBlue}{rgb}{0.53,0.81,0.98}
\definecolor{LightSlateBlue}{rgb}{0.52,0.44,1.00}
\definecolor{LightSlateGray}{rgb}{0.47,0.53,0.60}
\definecolor{LightSlateGrey}{rgb}{0.47,0.53,0.60}
\definecolor{LightSteelBlue1}{rgb}{0.79,0.88,1.00}
\definecolor{LightSteelBlue2}{rgb}{0.74,0.82,0.93}
\definecolor{LightSteelBlue3}{rgb}{0.64,0.71,0.80}
\definecolor{LightSteelBlue4}{rgb}{0.43,0.48,0.55}
\definecolor{LightSteelBlue}{rgb}{0.69,0.77,0.87}
\definecolor{LightYellow1}{rgb}{1.00,1.00,0.88}
\definecolor{LightYellow2}{rgb}{0.93,0.93,0.82}
\definecolor{LightYellow3}{rgb}{0.80,0.80,0.71}
\definecolor{LightYellow4}{rgb}{0.55,0.55,0.48}
\definecolor{LightYellow}{rgb}{1.00,1.00,0.88}
\definecolor{LimeGreen}{rgb}{0.20,0.80,0.20}
\definecolor{MediumAquamarine}{rgb}{0.40,0.80,0.67}
\definecolor{MediumBlue}{rgb}{0.00,0.00,0.80}
\definecolor{MediumOrchid1}{rgb}{0.88,0.40,1.00}
\definecolor{MediumOrchid2}{rgb}{0.82,0.37,0.93}
\definecolor{MediumOrchid3}{rgb}{0.71,0.32,0.80}
\definecolor{MediumOrchid4}{rgb}{0.48,0.22,0.55}
\definecolor{MediumOrchid}{rgb}{0.73,0.33,0.83}
\definecolor{MediumPurple1}{rgb}{0.67,0.51,1.00}
\definecolor{MediumPurple2}{rgb}{0.62,0.47,0.93}
\definecolor{MediumPurple3}{rgb}{0.54,0.41,0.80}
\definecolor{MediumPurple4}{rgb}{0.36,0.28,0.55}
\definecolor{MediumPurple}{rgb}{0.58,0.44,0.86}
\definecolor{MediumSeaGreen}{rgb}{0.24,0.70,0.44}
\definecolor{MediumSlateBlue}{rgb}{0.48,0.41,0.93}
\definecolor{MediumSpringGreen}{rgb}{0.00,0.98,0.60}
\definecolor{MediumTurquoise}{rgb}{0.28,0.82,0.80}
\definecolor{MediumVioletRed}{rgb}{0.78,0.08,0.52}
\definecolor{MidnightBlue}{rgb}{0.10,0.10,0.44}
\definecolor{MintCream}{rgb}{0.96,1.00,0.98}
\definecolor{MistyRose1}{rgb}{1.00,0.89,0.88}
\definecolor{MistyRose2}{rgb}{0.93,0.84,0.82}
\definecolor{MistyRose3}{rgb}{0.80,0.72,0.71}
\definecolor{MistyRose4}{rgb}{0.55,0.49,0.48}
\definecolor{MistyRose}{rgb}{1.00,0.89,0.88}
\definecolor{NavajoWhite1}{rgb}{1.00,0.87,0.68}
\definecolor{NavajoWhite2}{rgb}{0.93,0.81,0.63}
\definecolor{NavajoWhite3}{rgb}{0.80,0.70,0.55}
\definecolor{NavajoWhite4}{rgb}{0.55,0.47,0.37}
\definecolor{NavajoWhite}{rgb}{1.00,0.87,0.68}
\definecolor{NavyBlue}{rgb}{0.00,0.00,0.50}
\definecolor{OldLace}{rgb}{0.99,0.96,0.90}
\definecolor{OliveDrab1}{rgb}{0.75,1.00,0.24}
\definecolor{OliveDrab2}{rgb}{0.70,0.93,0.23}
\definecolor{OliveDrab3}{rgb}{0.60,0.80,0.20}
\definecolor{OliveDrab4}{rgb}{0.41,0.55,0.13}
\definecolor{OliveDrab}{rgb}{0.42,0.56,0.14}
\definecolor{OrangeRed1}{rgb}{1.00,0.27,0.00}
\definecolor{OrangeRed2}{rgb}{0.93,0.25,0.00}
\definecolor{OrangeRed3}{rgb}{0.80,0.22,0.00}
\definecolor{OrangeRed4}{rgb}{0.55,0.15,0.00}
\definecolor{OrangeRed}{rgb}{1.00,0.27,0.00}
\definecolor{PaleGoldenrod}{rgb}{0.93,0.91,0.67}
\definecolor{PaleGreen1}{rgb}{0.60,1.00,0.60}
\definecolor{PaleGreen2}{rgb}{0.56,0.93,0.56}
\definecolor{PaleGreen3}{rgb}{0.49,0.80,0.49}
\definecolor{PaleGreen4}{rgb}{0.33,0.55,0.33}
\definecolor{PaleGreen}{rgb}{0.60,0.98,0.60}
\definecolor{PaleTurquoise1}{rgb}{0.73,1.00,1.00}
\definecolor{PaleTurquoise2}{rgb}{0.68,0.93,0.93}
\definecolor{PaleTurquoise3}{rgb}{0.59,0.80,0.80}
\definecolor{PaleTurquoise4}{rgb}{0.40,0.55,0.55}
\definecolor{PaleTurquoise}{rgb}{0.69,0.93,0.93}
\definecolor{PaleVioletRed1}{rgb}{1.00,0.51,0.67}
\definecolor{PaleVioletRed2}{rgb}{0.93,0.47,0.62}
\definecolor{PaleVioletRed3}{rgb}{0.80,0.41,0.54}
\definecolor{PaleVioletRed4}{rgb}{0.55,0.28,0.36}
\definecolor{PaleVioletRed}{rgb}{0.86,0.44,0.58}
\definecolor{PapayaWhip}{rgb}{1.00,0.94,0.84}
\definecolor{PeachPuff1}{rgb}{1.00,0.85,0.73}
\definecolor{PeachPuff2}{rgb}{0.93,0.80,0.68}
\definecolor{PeachPuff3}{rgb}{0.80,0.69,0.58}
\definecolor{PeachPuff4}{rgb}{0.55,0.47,0.40}
\definecolor{PeachPuff}{rgb}{1.00,0.85,0.73}
\definecolor{PowderBlue}{rgb}{0.69,0.88,0.90}
\definecolor{RosyBrown1}{rgb}{1.00,0.76,0.76}
\definecolor{RosyBrown2}{rgb}{0.93,0.71,0.71}
\definecolor{RosyBrown3}{rgb}{0.80,0.61,0.61}
\definecolor{RosyBrown4}{rgb}{0.55,0.41,0.41}
\definecolor{RosyBrown}{rgb}{0.74,0.56,0.56}
\definecolor{RoyalBlue1}{rgb}{0.28,0.46,1.00}
\definecolor{RoyalBlue2}{rgb}{0.26,0.43,0.93}
\definecolor{RoyalBlue3}{rgb}{0.23,0.37,0.80}
\definecolor{RoyalBlue4}{rgb}{0.15,0.25,0.55}
\definecolor{RoyalBlue}{rgb}{0.25,0.41,0.88}
\definecolor{SaddleBrown}{rgb}{0.55,0.27,0.07}
\definecolor{SandyBrown}{rgb}{0.96,0.64,0.38}
\definecolor{SeaGreen1}{rgb}{0.33,1.00,0.62}
\definecolor{SeaGreen2}{rgb}{0.31,0.93,0.58}
\definecolor{SeaGreen3}{rgb}{0.26,0.80,0.50}
\definecolor{SeaGreen4}{rgb}{0.18,0.55,0.34}
\definecolor{SeaGreen}{rgb}{0.18,0.55,0.34}
\definecolor{SkyBlue1}{rgb}{0.53,0.81,1.00}
\definecolor{SkyBlue2}{rgb}{0.49,0.75,0.93}
\definecolor{SkyBlue3}{rgb}{0.42,0.65,0.80}
\definecolor{SkyBlue4}{rgb}{0.29,0.44,0.55}
\definecolor{SkyBlue}{rgb}{0.53,0.81,0.92}
\definecolor{SlateBlue1}{rgb}{0.51,0.44,1.00}
\definecolor{SlateBlue2}{rgb}{0.48,0.40,0.93}
\definecolor{SlateBlue3}{rgb}{0.41,0.35,0.80}
\definecolor{SlateBlue4}{rgb}{0.28,0.24,0.55}
\definecolor{SlateBlue}{rgb}{0.42,0.35,0.80}
\definecolor{SlateGray1}{rgb}{0.78,0.89,1.00}
\definecolor{SlateGray2}{rgb}{0.73,0.83,0.93}
\definecolor{SlateGray3}{rgb}{0.62,0.71,0.80}
\definecolor{SlateGray4}{rgb}{0.42,0.48,0.55}
\definecolor{SlateGray}{rgb}{0.44,0.50,0.56}
\definecolor{SlateGrey}{rgb}{0.44,0.50,0.56}
\definecolor{SpringGreen1}{rgb}{0.00,1.00,0.50}
\definecolor{SpringGreen2}{rgb}{0.00,0.93,0.46}
\definecolor{SpringGreen3}{rgb}{0.00,0.80,0.40}
\definecolor{SpringGreen4}{rgb}{0.00,0.55,0.27}
\definecolor{SpringGreen}{rgb}{0.00,1.00,0.50}
\definecolor{SteelBlue1}{rgb}{0.39,0.72,1.00}
\definecolor{SteelBlue2}{rgb}{0.36,0.67,0.93}
\definecolor{SteelBlue3}{rgb}{0.31,0.58,0.80}
\definecolor{SteelBlue4}{rgb}{0.21,0.39,0.55}
\definecolor{SteelBlue}{rgb}{0.27,0.51,0.71}
\definecolor{VioletRed1}{rgb}{1.00,0.24,0.59}
\definecolor{VioletRed2}{rgb}{0.93,0.23,0.55}
\definecolor{VioletRed3}{rgb}{0.80,0.20,0.47}
\definecolor{VioletRed4}{rgb}{0.55,0.13,0.32}
\definecolor{VioletRed}{rgb}{0.82,0.13,0.56}
\definecolor{WhiteSmoke}{rgb}{0.96,0.96,0.96}
\definecolor{YellowGreen}{rgb}{0.60,0.80,0.20}
\definecolor{aliceblue}{rgb}{0.94,0.97,1.00}
\definecolor{antiquewhite}{rgb}{0.98,0.92,0.84}
\definecolor{aquamarine1}{rgb}{0.50,1.00,0.83}
\definecolor{aquamarine2}{rgb}{0.46,0.93,0.78}
\definecolor{aquamarine3}{rgb}{0.40,0.80,0.67}
\definecolor{aquamarine4}{rgb}{0.27,0.55,0.45}
\definecolor{aquamarine}{rgb}{0.50,1.00,0.83}
\definecolor{azure1}{rgb}{0.94,1.00,1.00}
\definecolor{azure2}{rgb}{0.88,0.93,0.93}
\definecolor{azure3}{rgb}{0.76,0.80,0.80}
\definecolor{azure4}{rgb}{0.51,0.55,0.55}
\definecolor{azure}{rgb}{0.94,1.00,1.00}
\definecolor{beige}{rgb}{0.96,0.96,0.86}
\definecolor{bisque1}{rgb}{1.00,0.89,0.77}
\definecolor{bisque2}{rgb}{0.93,0.84,0.72}
\definecolor{bisque3}{rgb}{0.80,0.72,0.62}
\definecolor{bisque4}{rgb}{0.55,0.49,0.42}
\definecolor{bisque}{rgb}{1.00,0.89,0.77}
\definecolor{black}{rgb}{0.00,0.00,0.00}
\definecolor{blanchedalmond}{rgb}{1.00,0.92,0.80}
\definecolor{blue1}{rgb}{0.00,0.00,1.00}
\definecolor{blue2}{rgb}{0.00,0.00,0.93}
\definecolor{blue3}{rgb}{0.00,0.00,0.80}
\definecolor{blue4}{rgb}{0.00,0.00,0.55}
\definecolor{blueviolet}{rgb}{0.54,0.17,0.89}
\definecolor{blue}{rgb}{0.00,0.00,1.00}
\definecolor{brown1}{rgb}{1.00,0.25,0.25}
\definecolor{brown2}{rgb}{0.93,0.23,0.23}
\definecolor{brown3}{rgb}{0.80,0.20,0.20}
\definecolor{brown4}{rgb}{0.55,0.14,0.14}
\definecolor{brown}{rgb}{0.65,0.16,0.16}
\definecolor{burlywood1}{rgb}{1.00,0.83,0.61}
\definecolor{burlywood2}{rgb}{0.93,0.77,0.57}
\definecolor{burlywood3}{rgb}{0.80,0.67,0.49}
\definecolor{burlywood4}{rgb}{0.55,0.45,0.33}
\definecolor{burlywood}{rgb}{0.87,0.72,0.53}
\definecolor{cadetblue}{rgb}{0.37,0.62,0.63}
\definecolor{chartreuse1}{rgb}{0.50,1.00,0.00}
\definecolor{chartreuse2}{rgb}{0.46,0.93,0.00}
\definecolor{chartreuse3}{rgb}{0.40,0.80,0.00}
\definecolor{chartreuse4}{rgb}{0.27,0.55,0.00}
\definecolor{chartreuse}{rgb}{0.50,1.00,0.00}
\definecolor{chocolate1}{rgb}{1.00,0.50,0.14}
\definecolor{chocolate2}{rgb}{0.93,0.46,0.13}
\definecolor{chocolate3}{rgb}{0.80,0.40,0.11}
\definecolor{chocolate4}{rgb}{0.55,0.27,0.07}
\definecolor{chocolate}{rgb}{0.82,0.41,0.12}
\definecolor{coral1}{rgb}{1.00,0.45,0.34}
\definecolor{coral2}{rgb}{0.93,0.42,0.31}
\definecolor{coral3}{rgb}{0.80,0.36,0.27}
\definecolor{coral4}{rgb}{0.55,0.24,0.18}
\definecolor{coral}{rgb}{1.00,0.50,0.31}
\definecolor{cornflowerblue}{rgb}{0.39,0.58,0.93}
\definecolor{cornsilk1}{rgb}{1.00,0.97,0.86}
\definecolor{cornsilk2}{rgb}{0.93,0.91,0.80}
\definecolor{cornsilk3}{rgb}{0.80,0.78,0.69}
\definecolor{cornsilk4}{rgb}{0.55,0.53,0.47}
\definecolor{cornsilk}{rgb}{1.00,0.97,0.86}
\definecolor{cyan1}{rgb}{0.00,1.00,1.00}
\definecolor{cyan2}{rgb}{0.00,0.93,0.93}
\definecolor{cyan3}{rgb}{0.00,0.80,0.80}
\definecolor{cyan4}{rgb}{0.00,0.55,0.55}
\definecolor{cyan}{rgb}{0.00,1.00,1.00}
\definecolor{darkblue}{rgb}{0.00,0.00,0.55}
\definecolor{darkcyan}{rgb}{0.00,0.55,0.55}
\definecolor{darkgoldenrod}{rgb}{0.72,0.53,0.04}
\definecolor{darkgray}{rgb}{0.66,0.66,0.66}
\definecolor{darkgreen}{rgb}{0.00,0.39,0.00}
\definecolor{darkgrey}{rgb}{0.66,0.66,0.66}
\definecolor{darkkhaki}{rgb}{0.74,0.72,0.42}
\definecolor{darkmagenta}{rgb}{0.55,0.00,0.55}
\definecolor{darkolive}{rgb}{0.33,0.42,0.18}
\definecolor{darkorange}{rgb}{1.00,0.55,0.00}
\definecolor{darkorchid}{rgb}{0.60,0.20,0.80}
\definecolor{darkred}{rgb}{0.55,0.00,0.00}
\definecolor{darksalmon}{rgb}{0.91,0.59,0.48}
\definecolor{darksea}{rgb}{0.56,0.74,0.56}
\definecolor{darkslate}{rgb}{0.18,0.31,0.31}
\definecolor{darkslate}{rgb}{0.18,0.31,0.31}
\definecolor{darkslate}{rgb}{0.28,0.24,0.55}
\definecolor{darkturquoise}{rgb}{0.00,0.81,0.82}
\definecolor{darkviolet}{rgb}{0.58,0.00,0.83}
\definecolor{deeppink}{rgb}{1.00,0.08,0.58}
\definecolor{deepsky}{rgb}{0.00,0.75,1.00}
\definecolor{dimgray}{rgb}{0.41,0.41,0.41}
\definecolor{dimgrey}{rgb}{0.41,0.41,0.41}
\definecolor{dodgerblue}{rgb}{0.12,0.56,1.00}
\definecolor{firebrick1}{rgb}{1.00,0.19,0.19}
\definecolor{firebrick2}{rgb}{0.93,0.17,0.17}
\definecolor{firebrick3}{rgb}{0.80,0.15,0.15}
\definecolor{firebrick4}{rgb}{0.55,0.10,0.10}
\definecolor{firebrick}{rgb}{0.70,0.13,0.13}
\definecolor{floralwhite}{rgb}{1.00,0.98,0.94}
\definecolor{forestgreen}{rgb}{0.13,0.55,0.13}
\definecolor{gainsboro}{rgb}{0.86,0.86,0.86}
\definecolor{ghostwhite}{rgb}{0.97,0.97,1.00}
\definecolor{gold1}{rgb}{1.00,0.84,0.00}
\definecolor{gold2}{rgb}{0.93,0.79,0.00}
\definecolor{gold3}{rgb}{0.80,0.68,0.00}
\definecolor{gold4}{rgb}{0.55,0.46,0.00}
\definecolor{goldenrod1}{rgb}{1.00,0.76,0.15}
\definecolor{goldenrod2}{rgb}{0.93,0.71,0.13}
\definecolor{goldenrod3}{rgb}{0.80,0.61,0.11}
\definecolor{goldenrod4}{rgb}{0.55,0.41,0.08}
\definecolor{goldenrod}{rgb}{0.85,0.65,0.13}
\definecolor{gold}{rgb}{1.00,0.84,0.00}
\definecolor{gray0}{rgb}{0.00,0.00,0.00}
\definecolor{gray100}{rgb}{1.00,1.00,1.00}
\definecolor{gray10}{rgb}{0.10,0.10,0.10}
\definecolor{gray11}{rgb}{0.11,0.11,0.11}
\definecolor{gray12}{rgb}{0.12,0.12,0.12}
\definecolor{gray13}{rgb}{0.13,0.13,0.13}
\definecolor{gray14}{rgb}{0.14,0.14,0.14}
\definecolor{gray15}{rgb}{0.15,0.15,0.15}
\definecolor{gray16}{rgb}{0.16,0.16,0.16}
\definecolor{gray17}{rgb}{0.17,0.17,0.17}
\definecolor{gray18}{rgb}{0.18,0.18,0.18}
\definecolor{gray19}{rgb}{0.19,0.19,0.19}
\definecolor{gray1}{rgb}{0.01,0.01,0.01}
\definecolor{gray20}{rgb}{0.20,0.20,0.20}
\definecolor{gray21}{rgb}{0.21,0.21,0.21}
\definecolor{gray22}{rgb}{0.22,0.22,0.22}
\definecolor{gray23}{rgb}{0.23,0.23,0.23}
\definecolor{gray24}{rgb}{0.24,0.24,0.24}
\definecolor{gray25}{rgb}{0.25,0.25,0.25}
\definecolor{gray26}{rgb}{0.26,0.26,0.26}
\definecolor{gray27}{rgb}{0.27,0.27,0.27}
\definecolor{gray28}{rgb}{0.28,0.28,0.28}
\definecolor{gray29}{rgb}{0.29,0.29,0.29}
\definecolor{gray2}{rgb}{0.02,0.02,0.02}
\definecolor{gray30}{rgb}{0.30,0.30,0.30}
\definecolor{gray31}{rgb}{0.31,0.31,0.31}
\definecolor{gray32}{rgb}{0.32,0.32,0.32}
\definecolor{gray33}{rgb}{0.33,0.33,0.33}
\definecolor{gray34}{rgb}{0.34,0.34,0.34}
\definecolor{gray35}{rgb}{0.35,0.35,0.35}
\definecolor{gray36}{rgb}{0.36,0.36,0.36}
\definecolor{gray37}{rgb}{0.37,0.37,0.37}
\definecolor{gray38}{rgb}{0.38,0.38,0.38}
\definecolor{gray39}{rgb}{0.39,0.39,0.39}
\definecolor{gray3}{rgb}{0.03,0.03,0.03}
\definecolor{gray40}{rgb}{0.40,0.40,0.40}
\definecolor{gray41}{rgb}{0.41,0.41,0.41}
\definecolor{gray42}{rgb}{0.42,0.42,0.42}
\definecolor{gray43}{rgb}{0.43,0.43,0.43}
\definecolor{gray44}{rgb}{0.44,0.44,0.44}
\definecolor{gray45}{rgb}{0.45,0.45,0.45}
\definecolor{gray46}{rgb}{0.46,0.46,0.46}
\definecolor{gray47}{rgb}{0.47,0.47,0.47}
\definecolor{gray48}{rgb}{0.48,0.48,0.48}
\definecolor{gray49}{rgb}{0.49,0.49,0.49}
\definecolor{gray4}{rgb}{0.04,0.04,0.04}
\definecolor{gray50}{rgb}{0.50,0.50,0.50}
\definecolor{gray51}{rgb}{0.51,0.51,0.51}
\definecolor{gray52}{rgb}{0.52,0.52,0.52}
\definecolor{gray53}{rgb}{0.53,0.53,0.53}
\definecolor{gray54}{rgb}{0.54,0.54,0.54}
\definecolor{gray55}{rgb}{0.55,0.55,0.55}
\definecolor{gray56}{rgb}{0.56,0.56,0.56}
\definecolor{gray57}{rgb}{0.57,0.57,0.57}
\definecolor{gray58}{rgb}{0.58,0.58,0.58}
\definecolor{gray59}{rgb}{0.59,0.59,0.59}
\definecolor{gray5}{rgb}{0.05,0.05,0.05}
\definecolor{gray60}{rgb}{0.60,0.60,0.60}
\definecolor{gray61}{rgb}{0.61,0.61,0.61}
\definecolor{gray62}{rgb}{0.62,0.62,0.62}
\definecolor{gray63}{rgb}{0.63,0.63,0.63}
\definecolor{gray64}{rgb}{0.64,0.64,0.64}
\definecolor{gray65}{rgb}{0.65,0.65,0.65}
\definecolor{gray66}{rgb}{0.66,0.66,0.66}
\definecolor{gray67}{rgb}{0.67,0.67,0.67}
\definecolor{gray68}{rgb}{0.68,0.68,0.68}
\definecolor{gray69}{rgb}{0.69,0.69,0.69}
\definecolor{gray6}{rgb}{0.06,0.06,0.06}
\definecolor{gray70}{rgb}{0.70,0.70,0.70}
\definecolor{gray71}{rgb}{0.71,0.71,0.71}
\definecolor{gray72}{rgb}{0.72,0.72,0.72}
\definecolor{gray73}{rgb}{0.73,0.73,0.73}
\definecolor{gray74}{rgb}{0.74,0.74,0.74}
\definecolor{gray75}{rgb}{0.75,0.75,0.75}
\definecolor{gray76}{rgb}{0.76,0.76,0.76}
\definecolor{gray77}{rgb}{0.77,0.77,0.77}
\definecolor{gray78}{rgb}{0.78,0.78,0.78}
\definecolor{gray79}{rgb}{0.79,0.79,0.79}
\definecolor{gray7}{rgb}{0.07,0.07,0.07}
\definecolor{gray80}{rgb}{0.80,0.80,0.80}
\definecolor{gray81}{rgb}{0.81,0.81,0.81}
\definecolor{gray82}{rgb}{0.82,0.82,0.82}
\definecolor{gray83}{rgb}{0.83,0.83,0.83}
\definecolor{gray84}{rgb}{0.84,0.84,0.84}
\definecolor{gray85}{rgb}{0.85,0.85,0.85}
\definecolor{gray86}{rgb}{0.86,0.86,0.86}
\definecolor{gray87}{rgb}{0.87,0.87,0.87}
\definecolor{gray88}{rgb}{0.88,0.88,0.88}
\definecolor{gray89}{rgb}{0.89,0.89,0.89}
\definecolor{gray8}{rgb}{0.08,0.08,0.08}
\definecolor{gray90}{rgb}{0.90,0.90,0.90}
\definecolor{gray91}{rgb}{0.91,0.91,0.91}
\definecolor{gray92}{rgb}{0.92,0.92,0.92}
\definecolor{gray93}{rgb}{0.93,0.93,0.93}
\definecolor{gray94}{rgb}{0.94,0.94,0.94}
\definecolor{gray95}{rgb}{0.95,0.95,0.95}
\definecolor{gray96}{rgb}{0.96,0.96,0.96}
\definecolor{gray97}{rgb}{0.97,0.97,0.97}
\definecolor{gray98}{rgb}{0.98,0.98,0.98}
\definecolor{gray99}{rgb}{0.99,0.99,0.99}
\definecolor{gray9}{rgb}{0.09,0.09,0.09}
\definecolor{gray}{rgb}{0.75,0.75,0.75}
\definecolor{green1}{rgb}{0.00,1.00,0.00}
\definecolor{green2}{rgb}{0.00,0.93,0.00}
\definecolor{green3}{rgb}{0.00,0.80,0.00}
\definecolor{green4}{rgb}{0.00,0.55,0.00}
\definecolor{greenyellow}{rgb}{0.68,1.00,0.18}
\definecolor{green}{rgb}{0.00,1.00,0.00}
\definecolor{grey0}{rgb}{0.00,0.00,0.00}
\definecolor{grey100}{rgb}{1.00,1.00,1.00}
\definecolor{grey10}{rgb}{0.10,0.10,0.10}
\definecolor{grey11}{rgb}{0.11,0.11,0.11}
\definecolor{grey12}{rgb}{0.12,0.12,0.12}
\definecolor{grey13}{rgb}{0.13,0.13,0.13}
\definecolor{grey14}{rgb}{0.14,0.14,0.14}
\definecolor{grey15}{rgb}{0.15,0.15,0.15}
\definecolor{grey16}{rgb}{0.16,0.16,0.16}
\definecolor{grey17}{rgb}{0.17,0.17,0.17}
\definecolor{grey18}{rgb}{0.18,0.18,0.18}
\definecolor{grey19}{rgb}{0.19,0.19,0.19}
\definecolor{grey1}{rgb}{0.01,0.01,0.01}
\definecolor{grey20}{rgb}{0.20,0.20,0.20}
\definecolor{grey21}{rgb}{0.21,0.21,0.21}
\definecolor{grey22}{rgb}{0.22,0.22,0.22}
\definecolor{grey23}{rgb}{0.23,0.23,0.23}
\definecolor{grey24}{rgb}{0.24,0.24,0.24}
\definecolor{grey25}{rgb}{0.25,0.25,0.25}
\definecolor{grey26}{rgb}{0.26,0.26,0.26}
\definecolor{grey27}{rgb}{0.27,0.27,0.27}
\definecolor{grey28}{rgb}{0.28,0.28,0.28}
\definecolor{grey29}{rgb}{0.29,0.29,0.29}
\definecolor{grey2}{rgb}{0.02,0.02,0.02}
\definecolor{grey30}{rgb}{0.30,0.30,0.30}
\definecolor{grey31}{rgb}{0.31,0.31,0.31}
\definecolor{grey32}{rgb}{0.32,0.32,0.32}
\definecolor{grey33}{rgb}{0.33,0.33,0.33}
\definecolor{grey34}{rgb}{0.34,0.34,0.34}
\definecolor{grey35}{rgb}{0.35,0.35,0.35}
\definecolor{grey36}{rgb}{0.36,0.36,0.36}
\definecolor{grey37}{rgb}{0.37,0.37,0.37}
\definecolor{grey38}{rgb}{0.38,0.38,0.38}
\definecolor{grey39}{rgb}{0.39,0.39,0.39}
\definecolor{grey3}{rgb}{0.03,0.03,0.03}
\definecolor{grey40}{rgb}{0.40,0.40,0.40}
\definecolor{grey41}{rgb}{0.41,0.41,0.41}
\definecolor{grey42}{rgb}{0.42,0.42,0.42}
\definecolor{grey43}{rgb}{0.43,0.43,0.43}
\definecolor{grey44}{rgb}{0.44,0.44,0.44}
\definecolor{grey45}{rgb}{0.45,0.45,0.45}
\definecolor{grey46}{rgb}{0.46,0.46,0.46}
\definecolor{grey47}{rgb}{0.47,0.47,0.47}
\definecolor{grey48}{rgb}{0.48,0.48,0.48}
\definecolor{grey49}{rgb}{0.49,0.49,0.49}
\definecolor{grey4}{rgb}{0.04,0.04,0.04}
\definecolor{grey50}{rgb}{0.50,0.50,0.50}
\definecolor{grey51}{rgb}{0.51,0.51,0.51}
\definecolor{grey52}{rgb}{0.52,0.52,0.52}
\definecolor{grey53}{rgb}{0.53,0.53,0.53}
\definecolor{grey54}{rgb}{0.54,0.54,0.54}
\definecolor{grey55}{rgb}{0.55,0.55,0.55}
\definecolor{grey56}{rgb}{0.56,0.56,0.56}
\definecolor{grey57}{rgb}{0.57,0.57,0.57}
\definecolor{grey58}{rgb}{0.58,0.58,0.58}
\definecolor{grey59}{rgb}{0.59,0.59,0.59}
\definecolor{grey5}{rgb}{0.05,0.05,0.05}
\definecolor{grey60}{rgb}{0.60,0.60,0.60}
\definecolor{grey61}{rgb}{0.61,0.61,0.61}
\definecolor{grey62}{rgb}{0.62,0.62,0.62}
\definecolor{grey63}{rgb}{0.63,0.63,0.63}
\definecolor{grey64}{rgb}{0.64,0.64,0.64}
\definecolor{grey65}{rgb}{0.65,0.65,0.65}
\definecolor{grey66}{rgb}{0.66,0.66,0.66}
\definecolor{grey67}{rgb}{0.67,0.67,0.67}
\definecolor{grey68}{rgb}{0.68,0.68,0.68}
\definecolor{grey69}{rgb}{0.69,0.69,0.69}
\definecolor{grey6}{rgb}{0.06,0.06,0.06}
\definecolor{grey70}{rgb}{0.70,0.70,0.70}
\definecolor{grey71}{rgb}{0.71,0.71,0.71}
\definecolor{grey72}{rgb}{0.72,0.72,0.72}
\definecolor{grey73}{rgb}{0.73,0.73,0.73}
\definecolor{grey74}{rgb}{0.74,0.74,0.74}
\definecolor{grey75}{rgb}{0.75,0.75,0.75}
\definecolor{grey76}{rgb}{0.76,0.76,0.76}
\definecolor{grey77}{rgb}{0.77,0.77,0.77}
\definecolor{grey78}{rgb}{0.78,0.78,0.78}
\definecolor{grey79}{rgb}{0.79,0.79,0.79}
\definecolor{grey7}{rgb}{0.07,0.07,0.07}
\definecolor{grey80}{rgb}{0.80,0.80,0.80}
\definecolor{grey81}{rgb}{0.81,0.81,0.81}
\definecolor{grey82}{rgb}{0.82,0.82,0.82}
\definecolor{grey83}{rgb}{0.83,0.83,0.83}
\definecolor{grey84}{rgb}{0.84,0.84,0.84}
\definecolor{grey85}{rgb}{0.85,0.85,0.85}
\definecolor{grey86}{rgb}{0.86,0.86,0.86}
\definecolor{grey87}{rgb}{0.87,0.87,0.87}
\definecolor{grey88}{rgb}{0.88,0.88,0.88}
\definecolor{grey89}{rgb}{0.89,0.89,0.89}
\definecolor{grey8}{rgb}{0.08,0.08,0.08}
\definecolor{grey90}{rgb}{0.90,0.90,0.90}
\definecolor{grey91}{rgb}{0.91,0.91,0.91}
\definecolor{grey92}{rgb}{0.92,0.92,0.92}
\definecolor{grey93}{rgb}{0.93,0.93,0.93}
\definecolor{grey94}{rgb}{0.94,0.94,0.94}
\definecolor{grey95}{rgb}{0.95,0.95,0.95}
\definecolor{grey96}{rgb}{0.96,0.96,0.96}
\definecolor{grey97}{rgb}{0.97,0.97,0.97}
\definecolor{grey98}{rgb}{0.98,0.98,0.98}
\definecolor{grey99}{rgb}{0.99,0.99,0.99}
\definecolor{grey9}{rgb}{0.09,0.09,0.09}
\definecolor{grey}{rgb}{0.75,0.75,0.75}
\definecolor{honeydew1}{rgb}{0.94,1.00,0.94}
\definecolor{honeydew2}{rgb}{0.88,0.93,0.88}
\definecolor{honeydew3}{rgb}{0.76,0.80,0.76}
\definecolor{honeydew4}{rgb}{0.51,0.55,0.51}
\definecolor{honeydew}{rgb}{0.94,1.00,0.94}
\definecolor{hotpink}{rgb}{1.00,0.41,0.71}
\definecolor{indianred}{rgb}{0.80,0.36,0.36}
\definecolor{ivory1}{rgb}{1.00,1.00,0.94}
\definecolor{ivory2}{rgb}{0.93,0.93,0.88}
\definecolor{ivory3}{rgb}{0.80,0.80,0.76}
\definecolor{ivory4}{rgb}{0.55,0.55,0.51}
\definecolor{ivory}{rgb}{1.00,1.00,0.94}
\definecolor{khaki1}{rgb}{1.00,0.96,0.56}
\definecolor{khaki2}{rgb}{0.93,0.90,0.52}
\definecolor{khaki3}{rgb}{0.80,0.78,0.45}
\definecolor{khaki4}{rgb}{0.55,0.53,0.31}
\definecolor{khaki}{rgb}{0.94,0.90,0.55}
\definecolor{lavenderblush}{rgb}{1.00,0.94,0.96}
\definecolor{lavender}{rgb}{0.90,0.90,0.98}
\definecolor{lawngreen}{rgb}{0.49,0.99,0.00}
\definecolor{lemonchiffon}{rgb}{1.00,0.98,0.80}
\definecolor{lightblue}{rgb}{0.68,0.85,0.90}
\definecolor{lightcoral}{rgb}{0.94,0.50,0.50}
\definecolor{lightcyan}{rgb}{0.88,1.00,1.00}
\definecolor{lightgoldenrod}{rgb}{0.93,0.87,0.51}
\definecolor{lightgoldenrod}{rgb}{0.98,0.98,0.82}
\definecolor{lightgray}{rgb}{0.83,0.83,0.83}
\definecolor{lightgreen}{rgb}{0.56,0.93,0.56}
\definecolor{lightgrey}{rgb}{0.83,0.83,0.83}
\definecolor{lightpink}{rgb}{1.00,0.71,0.76}
\definecolor{lightsalmon}{rgb}{1.00,0.63,0.48}
\definecolor{lightsea}{rgb}{0.13,0.70,0.67}
\definecolor{lightsky}{rgb}{0.53,0.81,0.98}
\definecolor{lightslate}{rgb}{0.47,0.53,0.60}
\definecolor{lightslate}{rgb}{0.47,0.53,0.60}
\definecolor{lightslate}{rgb}{0.52,0.44,1.00}
\definecolor{lightsteel}{rgb}{0.69,0.77,0.87}
\definecolor{lightyellow}{rgb}{1.00,1.00,0.88}
\definecolor{limegreen}{rgb}{0.20,0.80,0.20}
\definecolor{linen}{rgb}{0.98,0.94,0.90}
\definecolor{magenta1}{rgb}{1.00,0.00,1.00}
\definecolor{magenta2}{rgb}{0.93,0.00,0.93}
\definecolor{magenta3}{rgb}{0.80,0.00,0.80}
\definecolor{magenta4}{rgb}{0.55,0.00,0.55}
\definecolor{magenta}{rgb}{1.00,0.00,1.00}
\definecolor{maroon1}{rgb}{1.00,0.20,0.70}
\definecolor{maroon2}{rgb}{0.93,0.19,0.65}
\definecolor{maroon3}{rgb}{0.80,0.16,0.56}
\definecolor{maroon4}{rgb}{0.55,0.11,0.38}
\definecolor{maroon}{rgb}{0.69,0.19,0.38}
\definecolor{mediumaquamarine}{rgb}{0.40,0.80,0.67}
\definecolor{mediumblue}{rgb}{0.00,0.00,0.80}
\definecolor{mediumorchid}{rgb}{0.73,0.33,0.83}
\definecolor{mediumpurple}{rgb}{0.58,0.44,0.86}
\definecolor{mediumsea}{rgb}{0.24,0.70,0.44}
\definecolor{mediumslate}{rgb}{0.48,0.41,0.93}
\definecolor{mediumspring}{rgb}{0.00,0.98,0.60}
\definecolor{mediumturquoise}{rgb}{0.28,0.82,0.80}
\definecolor{mediumviolet}{rgb}{0.78,0.08,0.52}
\definecolor{midnightblue}{rgb}{0.10,0.10,0.44}
\definecolor{mintcream}{rgb}{0.96,1.00,0.98}
\definecolor{mistyrose}{rgb}{1.00,0.89,0.88}
\definecolor{moccasin}{rgb}{1.00,0.89,0.71}
\definecolor{navajowhite}{rgb}{1.00,0.87,0.68}
\definecolor{navyblue}{rgb}{0.00,0.00,0.50}
\definecolor{navy}{rgb}{0.00,0.00,0.50}
\definecolor{oldlace}{rgb}{0.99,0.96,0.90}
\definecolor{olivedrab}{rgb}{0.42,0.56,0.14}
\definecolor{orange1}{rgb}{1.00,0.65,0.00}
\definecolor{orange2}{rgb}{0.93,0.60,0.00}
\definecolor{orange3}{rgb}{0.80,0.52,0.00}
\definecolor{orange4}{rgb}{0.55,0.35,0.00}
\definecolor{orangered}{rgb}{1.00,0.27,0.00}
\definecolor{orange}{rgb}{1.00,0.65,0.00}
\definecolor{orchid1}{rgb}{1.00,0.51,0.98}
\definecolor{orchid2}{rgb}{0.93,0.48,0.91}
\definecolor{orchid3}{rgb}{0.80,0.41,0.79}
\definecolor{orchid4}{rgb}{0.55,0.28,0.54}
\definecolor{orchid}{rgb}{0.85,0.44,0.84}
\definecolor{palegoldenrod}{rgb}{0.93,0.91,0.67}
\definecolor{palegreen}{rgb}{0.60,0.98,0.60}
\definecolor{paleturquoise}{rgb}{0.69,0.93,0.93}
\definecolor{paleviolet}{rgb}{0.86,0.44,0.58}
\definecolor{papayawhip}{rgb}{1.00,0.94,0.84}
\definecolor{peachpuff}{rgb}{1.00,0.85,0.73}
\definecolor{peru}{rgb}{0.80,0.52,0.25}
\definecolor{pink1}{rgb}{1.00,0.71,0.77}
\definecolor{pink2}{rgb}{0.93,0.66,0.72}
\definecolor{pink3}{rgb}{0.80,0.57,0.62}
\definecolor{pink4}{rgb}{0.55,0.39,0.42}
\definecolor{pink}{rgb}{1.00,0.75,0.80}
\definecolor{plum1}{rgb}{1.00,0.73,1.00}
\definecolor{plum2}{rgb}{0.93,0.68,0.93}
\definecolor{plum3}{rgb}{0.80,0.59,0.80}
\definecolor{plum4}{rgb}{0.55,0.40,0.55}
\definecolor{plum}{rgb}{0.87,0.63,0.87}
\definecolor{powderblue}{rgb}{0.69,0.88,0.90}
\definecolor{purple1}{rgb}{0.61,0.19,1.00}
\definecolor{purple2}{rgb}{0.57,0.17,0.93}
\definecolor{purple3}{rgb}{0.49,0.15,0.80}
\definecolor{purple4}{rgb}{0.33,0.10,0.55}
\definecolor{purple}{rgb}{0.63,0.13,0.94}
\definecolor{red1}{rgb}{1.00,0.00,0.00}
\definecolor{red2}{rgb}{0.93,0.00,0.00}
\definecolor{red3}{rgb}{0.80,0.00,0.00}
\definecolor{red4}{rgb}{0.55,0.00,0.00}
\definecolor{red}{rgb}{1.00,0.00,0.00}
\definecolor{rosybrown}{rgb}{0.74,0.56,0.56}
\definecolor{royalblue}{rgb}{0.25,0.41,0.88}
\definecolor{saddlebrown}{rgb}{0.55,0.27,0.07}
\definecolor{salmon1}{rgb}{1.00,0.55,0.41}
\definecolor{salmon2}{rgb}{0.93,0.51,0.38}
\definecolor{salmon3}{rgb}{0.80,0.44,0.33}
\definecolor{salmon4}{rgb}{0.55,0.30,0.22}
\definecolor{salmon}{rgb}{0.98,0.50,0.45}
\definecolor{sandybrown}{rgb}{0.96,0.64,0.38}
\definecolor{seagreen}{rgb}{0.18,0.55,0.34}
\definecolor{seashell1}{rgb}{1.00,0.96,0.93}
\definecolor{seashell2}{rgb}{0.93,0.90,0.87}
\definecolor{seashell3}{rgb}{0.80,0.77,0.75}
\definecolor{seashell4}{rgb}{0.55,0.53,0.51}
\definecolor{seashell}{rgb}{1.00,0.96,0.93}
\definecolor{sienna1}{rgb}{1.00,0.51,0.28}
\definecolor{sienna2}{rgb}{0.93,0.47,0.26}
\definecolor{sienna3}{rgb}{0.80,0.41,0.22}
\definecolor{sienna4}{rgb}{0.55,0.28,0.15}
\definecolor{sienna}{rgb}{0.63,0.32,0.18}
\definecolor{skyblue}{rgb}{0.53,0.81,0.92}
\definecolor{slateblue}{rgb}{0.42,0.35,0.80}
\definecolor{slategray}{rgb}{0.44,0.50,0.56}
\definecolor{slategrey}{rgb}{0.44,0.50,0.56}
\definecolor{snow1}{rgb}{1.00,0.98,0.98}
\definecolor{snow2}{rgb}{0.93,0.91,0.91}
\definecolor{snow3}{rgb}{0.80,0.79,0.79}
\definecolor{snow4}{rgb}{0.55,0.54,0.54}
\definecolor{snow}{rgb}{1.00,0.98,0.98}
\definecolor{springgreen}{rgb}{0.00,1.00,0.50}
\definecolor{steelblue}{rgb}{0.27,0.51,0.71}
\definecolor{tan1}{rgb}{1.00,0.65,0.31}
\definecolor{tan2}{rgb}{0.93,0.60,0.29}
\definecolor{tan3}{rgb}{0.80,0.52,0.25}
\definecolor{tan4}{rgb}{0.55,0.35,0.17}
\definecolor{tan}{rgb}{0.82,0.71,0.55}
\definecolor{thistle1}{rgb}{1.00,0.88,1.00}
\definecolor{thistle2}{rgb}{0.93,0.82,0.93}
\definecolor{thistle3}{rgb}{0.80,0.71,0.80}
\definecolor{thistle4}{rgb}{0.55,0.48,0.55}
\definecolor{thistle}{rgb}{0.85,0.75,0.85}
\definecolor{tomato1}{rgb}{1.00,0.39,0.28}
\definecolor{tomato2}{rgb}{0.93,0.36,0.26}
\definecolor{tomato3}{rgb}{0.80,0.31,0.22}
\definecolor{tomato4}{rgb}{0.55,0.21,0.15}
\definecolor{tomato}{rgb}{1.00,0.39,0.28}
\definecolor{turquoise1}{rgb}{0.00,0.96,1.00}
\definecolor{turquoise2}{rgb}{0.00,0.90,0.93}
\definecolor{turquoise3}{rgb}{0.00,0.77,0.80}
\definecolor{turquoise4}{rgb}{0.00,0.53,0.55}
\definecolor{turquoise}{rgb}{0.25,0.88,0.82}
\definecolor{violetred}{rgb}{0.82,0.13,0.56}
\definecolor{violet}{rgb}{0.93,0.51,0.93}
\definecolor{wheat1}{rgb}{1.00,0.91,0.73}
\definecolor{wheat2}{rgb}{0.93,0.85,0.68}
\definecolor{wheat3}{rgb}{0.80,0.73,0.59}
\definecolor{wheat4}{rgb}{0.55,0.49,0.40}
\definecolor{wheat}{rgb}{0.96,0.87,0.70}
\definecolor{whitesmoke}{rgb}{0.96,0.96,0.96}
\definecolor{white}{rgb}{1.00,1.00,1.00}
\definecolor{yellow1}{rgb}{1.00,1.00,0.00}
\definecolor{yellow2}{rgb}{0.93,0.93,0.00}
\definecolor{yellow3}{rgb}{0.80,0.80,0.00}
\definecolor{yellow4}{rgb}{0.55,0.55,0.00}
\definecolor{yellowgreen}{rgb}{0.60,0.80,0.20}
\definecolor{yellow}{rgb}{1.00,1.00,0.00}